\pdfoutput=1
\documentclass[prl,reprint, epsfig,amssymb,bibliography,amsmath,footinbib]{revtex4-2}

\usepackage{blkarray}

\usepackage{amssymb,amsmath,bm,epsfig,graphicx,subfigure,hyperref}

\usepackage[utf8]{inputenc}
\usepackage[vietnamese, english]{babel}
\usepackage{chngcntr}
\usepackage{xcolor}
\usepackage{braket}
\usepackage{physics}
\usepackage{sidecap}
\usepackage{lipsum}
\usepackage{amsfonts}
\usepackage{amsmath, amssymb}

\begin{document}

\title{Quantum Geometric Oscillations in Two-Dimensional Flat-Band Solids}
\author{\foreignlanguage{vietnamese}{Võ Tiến Phong}}
\affiliation{Department of Physics and Astronomy, University of Pennsylvania, Philadelphia, PA 19104}
\author{E. J. Mele}
\affiliation{Department of Physics and Astronomy, University of Pennsylvania, Philadelphia, PA 19104}
\date{\today}

\begin{abstract} 
Two-dimensional van der Waals heterostructures can be engineered into artificial superlattices that host flat bands with significant Berry curvature and provide a favorable environment for the emergence of novel electron dynamics. In particular, the Berry curvature can induce an oscillating trajectory of an electron wave packet transverse to an applied static electric field. Though analogous to Bloch oscillations, this novel oscillatory behavior is driven entirely by quantum geometry in momentum space instead of band dispersion. While the current from Bloch oscillations can be localized by increasing field strength, the current from the geometric orbits saturates to a nonzero plateau in the strong-field limit. In nonmagnetic materials, the geometric oscillations are even under inversion of the applied field, whereas the Bloch oscillations are odd, a property that can be used to distinguish these two coexisting effects.
\end{abstract}

\maketitle

Stacking or patterning atomically thin two-dimensional materials can produce translationally ordered artificial crystals with  superlattice periods that greatly exceed the  atomic scale \cite{ T05, AM20, MS22}. This introduces two new generic features into their electronic behavior. First, the electronic energy bands fracture into spectrally isolated miniband manifolds that disperse weakly across a compressed Brillouin zone \cite{MCVPB10,BM11,MK12,ZWWT20,LLBdJ21}. It was recognized  decades ago that this feature provides a favorable environment for  Bloch oscillations in quasi-one-dimensional semiconductor superlattices \cite{LBBSK92, FLSM92, DOKK95, BL95, ISGRD95}. These Bloch oscillations originate  from band dispersion, and are a prototypical signature of quantum electron dynamics whereby a static applied electric field induces an oscillating current \cite{W60, MB93}. Inspired by  recent progress in twistronic research, it has  been suggested that analogous two-dimensional Bloch oscillations might be observable in  moiré materials as well \cite{ FDKL21, VOB22}. Second, moiré minibands can be manipulated to produce momentum-space band inversions where the orbital character of neighboring multiplets exchanges \cite{WLT19,GZX20, PLG21, DCZF21,ZXZ22}. This introduces a fundamentally different type of oscillations driven by quantum geometry encoded in the momentum-space Berry curvature with important consequences for transport. Unlike Bloch oscillations, these \textit{geometric oscillations} do not exist in one dimension, and are especially relevant to modern topological moiré materials.

In this Letter, we elucidate the characteristics of these geometric oscillations in non-magnetic crystals both in the semiclassical description and in the Wannier-Stark framework. We find that the  frequencies for both Bloch and geometric oscillations share the same dependence on the applied field. However, the sizes and shapes of the orbits are drastically different in the two cases. For the conventional Bloch oscillations, the  direction depends sensitively on the orientation of the applied field relative to the symmetry axes of the crystal, and the  amplitude decays inversely proportionally to the field magnitude. On the other hand, the geometric oscillations always propagate transverse to the applied field, and their amplitudes are characteristically independent of the electric field magnitude. In ultracold optical lattices, the geometric influence on wavepacket dynamics has already been used to map the Berry curvature \cite{PC12, WPCP17}. However, in solid-state media, directly measuring wavepacket dynamics is difficult; thus, we focus on transport signatures instead. In transport, the incompressibility of geometric orbits translates to a generic finite residual drift Hall current at strong fields, in stark contrast to the vanishing drift current originating from Bloch oscillations due to Wannier-Stark localization \cite{EH87, K90,PB91, LM91, N91}. Thus, the diagnostic of Bloch oscillations via measuring negative differential conductance  breaks down for the geometric oscillations \cite{ET70}. Instead, the signature of geometric oscillations is a saturation of the drift Hall currents as field strength increases. Furthermore, the drift currents from Bloch oscillations change sign when the applied field is inverted while the drift currents from geometric oscillations  are invariant under field inversion, a property that allows these two currents to be distinguished when they coexist. This property was also noted in Ref. \cite{PC12}. In addition to being a fascinating demonstration of the quantum geometric nature of electrons in crystalline solids, observing these geometric oscillations would be an important step towards the ambitious goal of eventually deploying these oscillatory currents as radiation sources.

We start with the dynamics of a wave packet in an isolated $n^\text{th}$ band  governed by the  semiclassical equations \cite{CN95,CN96,XCN10}
\begin{equation}
    \begin{split}
    \label{eq: semiclassical dynamics}
         \hbar\dot{\mathbf{k}} &= -e \mathbf{E}, \quad \hbar\dot{\mathbf{r}}_n =  \nabla_\mathbf{k}\varepsilon_{n,\mathbf{k}}  + \hbar\dot{\mathbf{k}} \times \boldsymbol{\Omega}_{n,\mathbf{k}},
    \end{split}
\end{equation}
where $\varepsilon_{n,\mathbf{k}}$ and $\boldsymbol{\Omega}_{n,\mathbf{k}} = \Omega_{n,\mathbf{k}} \hat{\mathbf{e}}_\mathbf{z}$ are the band dispersion and Berry curvature. Both of these functions are real and $\mathbf{k}$-periodic, and thus,  can be expanded in Fourier series $\varepsilon_{n,\mathbf{k}} = \sum_{\mathbf{a}_i} e^{i \mathbf{a}_i \cdot \mathbf{k}} \tilde{\varepsilon}_{n,\mathbf{a}_i}$ and $ \Omega_{n,\mathbf{k}} = \sum_{\mathbf{a}_i} e^{i \mathbf{a}_i \cdot \mathbf{k}} \tilde{\Omega}_{n,\mathbf{a}_i},$ and $\mathbf{a}_i$ is a lattice translation vector. Integrating $\dot{\mathbf{r}}_n$ with respect to $t,$ we obtain three distinct contributions: 
\begin{equation}
    \begin{split}
        \mathbf{r}_\text{drift}(t) & = \frac{t}{\hbar}\sum_{\mathbf{a}_i\cdot \mathbf{E} = 0}  e^{i \mathbf{a}_i \cdot \mathbf{k}_0} \left[ i \mathbf{a}_i\tilde{\varepsilon}_{n,\mathbf{a}_i} - e  \mathbf{E} \times\tilde{\boldsymbol{\Omega}}_{n,\mathbf{a}_i} \right], \\ 
        \mathbf{r}_\text{Bloch}(t) & = \sum_{\mathbf{a}_i\cdot \mathbf{E} \neq 0} e^{i \mathbf{a}_i \cdot \mathbf{k}_0}\left[1-e^{-i e \mathbf{a}_i \cdot \mathbf{E}t/\hbar } \right]   \frac{\mathbf{a}_i\tilde{\varepsilon}_{n,\mathbf{a}_i}}{e \mathbf{a}_i \cdot \mathbf{E}},   \\
        \mathbf{r}_\text{geom}(t) & = i\sum_{\mathbf{a}_i \cdot \mathbf{E}\neq 0} e^{i \mathbf{a}_i \cdot \mathbf{k}_0}\left[1-e^{-i e \mathbf{a}_i \cdot \mathbf{E}t/\hbar } \right] \frac{ \mathbf{E} \times\tilde{\boldsymbol{\Omega}}_{n,\mathbf{a}_i}}{\mathbf{a}_i \cdot \mathbf{E}}.
    \end{split}
\end{equation}
The drift term includes a topological contribution that is directly proportional to the Chern number $\mathcal{C} = \int \frac{d^2 \mathbf{k}}{2\pi} \Omega_n(\mathbf{k}) \propto \tilde{\Omega}_{n,\mathbf{0}}.$ Thus, it makes sense that this term is a constant drift instead of an oscillation because it transports electrons from one edge to another. All the other terms in $\mathbf{r}_\text{drift}(t)$ occur along lines where the electric field is perpendicular to a lattice vector $\mathbf{a}_i.$ This is because these components experience no acceleration due to the electric field, and thus their contributions to the velocity are constant. The Bloch oscillating term $\mathbf{r}_\text{Bloch}(t)$ is caused by the usual band dispersion. In general, it has a complicated directional dependence. The crystal momentum monotonically traces out a straight path in $\mathbf{k}$ along the direction of $\mathbf{E}.$ At each $\mathbf{k}$ point, the velocity points along the direction of the gradient of the dispersion. The amplitude of these orbits  diverges as $E^{-1}.$ That is, at weak field strengths, the amplitude is macroscopic, much larger than the mean free path. So a scattering event likely occurs before an electron had time to execute a complete orbit, making Bloch oscillations notoriously difficult to observe. In contrast, the geometric term $\mathbf{r}_\text{geom}(t)$ has a simple directional dependence, pointing always orthogonally to $\mathbf{E}.$ The amplitude of this term is characteristically \textit{independent} of the magnitude of $\mathbf{E}.$ Meanwhile, the oscillation frequency increases with $E.$ In what follows, we will ignore the drift contributions, and focus entirely on the oscillatory behavior.

 The consequences of geometric oscillations can be interrogated from a complementary perspective using the Wannier-Stark formalism \cite{W60,S72,GH98,HKKM04,WKKM04}, which offers real-space insights absent in the semiclassical description.  We begin with a tight-binding lattice that contains $\sigma$ atomic orbitals located at $\boldsymbol{\tau}_\sigma$ within a unit cell, denoted by  $\ket{A_{\mathbf{a}_i,\sigma}}$, and an electric field $\mathbf{E}$. These orbitals are assumed to be site localized such that the position operator $\hat{\mathbf{r}}$ is diagonal in this basis. In momentum space with $\ket{\varphi_{\mathbf{k},\sigma}} = \mathcal{N}^{-\frac{1}{2}} \sum_{\mathbf{a}_i} \ket{A_{\mathbf{a}_i,\sigma}} e^{i \mathbf{k} \cdot \left( \mathbf{a}_i + \boldsymbol{\tau}_\sigma \right)},$ where $\mathcal{N}$ is the number of unit cells, the Hamiltonian is \footnote{We should point out that in order for the derivative with respect to $\mathbf{k}$ to be defined, we must regard $\mathbf{k}$ to be a continuous variable, which corresponds to the limit of $\mathcal{N} \rightarrow \infty.$ Throughout this work, we assume this limit implicitly; thus, we use $\mathcal{N}^{-1} \sum_{\mathbf{k}}$ and $V_\text{BZ}^{-1}\int d^2 \mathbf{k}$ interchangeably.}
\begin{equation}
\label{eq: Hamiltonian in k space}
    \hat{\mathcal{H}} = \sum_{\mathbf{k},\sigma',\sigma} \ket{\varphi_{\mathbf{k},\sigma'}} \left[ \mathcal{H}_0^{\sigma',\sigma}(\mathbf{k}) + i \delta_{\sigma',\sigma} e \mathbf{E} \cdot \nabla_\mathbf{k} \right] \bra{\varphi_{\mathbf{k},\sigma}},
\end{equation}
where $\mathcal{H}_0^{\sigma',\sigma}(\mathbf{k}) = \sum_{\mathbf{a}_i} e^{i \mathbf{k} \cdot \left( \mathbf{a}_i + \boldsymbol{\tau}_{\sigma}- \boldsymbol{\tau}_{\sigma'} \right)} \bra{A_{\mathbf{0},\sigma'}} \hat{\mathcal{H}}_0 \ket{A_{\mathbf{a}_i,\sigma}}$ and $\hat{\mathcal{H}}_0$ is the Hamiltonian without electric field. The gradient acts to the right.  We now project this Hamiltonian to an \textit{isolated} energy band of interest with eigenstates $\ket{\psi_{\mathbf{k}}}$ and energies $\varepsilon_\mathbf{k}$ satisfying $\mathcal{H}_0(\mathbf{k})\chi_{\mathbf{k}} = \varepsilon_\mathbf{k} \chi_{\mathbf{k}}$ and $\ket{\psi_\mathbf{k}} = \sum_{\sigma} \chi_{\mathbf{k},\sigma} \ket{\varphi_{\mathbf{k},\sigma}}.$ The band index is  implicit. This band projection is permissible when there at least exists sizable gaps between the band of interest and other energy bands so that Zener tunneling can be neglected \cite{Z32}. We find \cite{K61,FBF73,DVP94}
\begin{equation}
\label{eq: effective Hamiltonian}
    \hat{\mathcal{H}}_\text{eff} =  \sum_{\mathbf{k}} \ket{\psi_{\mathbf{k}}} \left[\varepsilon_\mathbf{k} + e \mathbf{E} \cdot \boldsymbol{\mathcal{A}}_\mathbf{k}+ie \mathbf{E} \cdot \nabla_\mathbf{k} \right] \bra{\psi_\mathbf{k}},
\end{equation}
where $\boldsymbol{\mathcal{A}}_\mathbf{k} = i \chi_\mathbf{k}^\dagger \nabla_\mathbf{k} \chi_\mathbf{k}$ is the Berry connection. For the Berry connection to be well-defined, $\chi_\mathbf{k}$ must be differentiable, which can always be chosen for Chern-trivial bands; we assume this throughout since the absence of a Chern number actually makes the oscillatory behavior clearer. Also, although the Berry connection is not gauge-invariant,  we will demonstrate that the results are indeed gauge-invariant.

The single-band effective Hamiltonian in Eq. \eqref{eq: effective Hamiltonian} can be diagonalized exactly by solving the partial differential equation $\mathcal{E}\Psi_\mathbf{k} = \left[\varepsilon_\mathbf{k} + e \mathbf{E} \cdot \boldsymbol{\mathcal{A}}_\mathbf{k}+ie \mathbf{E} \cdot \nabla_\mathbf{k} \right] \Psi_\mathbf{k}.$ Requiring $\Psi_{\mathbf{k}} = \Psi_{\mathbf{k} + \mathbf{G}},$ we can write $\Psi_\mathbf{k} = \sum_{\mathbf{a}_i} e^{i \mathbf{k} \cdot \mathbf{a}_i} \tilde{\Psi}_{\mathbf{a}_i}$ and find that the eigenvalue problem can be cast in the form of a matrix equation
\begin{equation}
    \sum_{\mathbf{a}_i'} \left[\tilde{\varepsilon}_{\mathbf{a}_i-\mathbf{a}_i'} + e \mathbf{E} \cdot \tilde{\boldsymbol{\mathcal{A}}}_{\mathbf{a}_i-\mathbf{a}_i'} \right]\tilde{\Psi}_{\mathbf{a}_i'}- e\mathbf{E}\cdot \mathbf{a}_i \tilde{\Psi}_{\mathbf{a}_i} = \mathcal{E} \tilde{\Psi}_{\mathbf{a}_i}.
\end{equation}
This is an  $\mathcal{N}\times\mathcal{N}$ matrix equation with $\mathcal{N}$ eigenvalues. If $\Psi_\mathbf{k}$ is an eigenstate with eigenvalue $\mathcal{E},$ then $\Psi_\mathbf{k}e^{-i\mathbf{k}\cdot \mathbf{a}_i}$ is also an eigenstate, not necessarily independent, with eigenvalue $\mathcal{E}+e\mathbf{E}\cdot \mathbf{a}_i.$ When $\mathbf{E} \cdot \mathbf{a}_i \neq 0$ for any nonzero $\mathbf{a}_i,$ then a complete orthonormal set of solutions can be written exactly as $\bra{\psi_\mathbf{k}}\ket{\Psi_{\mathbf{a}_i}} =\Psi_{\mathbf{k},\mathbf{a}_i} = \mathcal{N}^{-\frac{1}{2}} \exp \left(i \theta_\mathbf{k} - i \mathbf{k}\cdot \mathbf{a}_i \right)$ with energies $\mathcal{E}_{\mathbf{a}_i} = e\mathbf{E} \cdot \mathbf{a}_i+\tilde{\varepsilon}_{\mathbf{0}} + e \mathbf{E} \cdot \boldsymbol{\tilde{\mathcal{A}}}_{\mathbf{0}},$ where $\theta_\mathbf{k} = \sum_{\mathbf{a}_i \neq \mathbf{0}} \left[\tilde{\varepsilon}_{\mathbf{a}_i} + e \mathbf{E} \cdot \tilde{\boldsymbol{\mathcal{A}}}_{\mathbf{a}_i}\right]\left[ie \mathbf{E}\cdot \mathbf{a}_i\right]^{-1} e^{i \mathbf{k} \cdot \mathbf{a}_i}.$ It is clear that $\theta_\mathbf{k}$ is a real function. We observe that the energy levels form a discrete ladder with spacing given by $e\mathbf{a}_i\cdot \mathbf{E};$ this is the two-dimensional counterpart to the usual Wannier-Stark ladder \cite{W60}. The Berry connection plays no role in the relative spacing between ladder rungs; it only affects the zero of energy, which we can shift so that $\mathbf{a}_i = \mathbf{0},$ $\mathcal{E}_\mathbf{0} = 0.$

The wavefunction $\ket{\Psi_{\mathbf{a}_i}}$ is centered inside unit cell $\mathbf{a}_i$ with spatial spread $\langle \delta\hat{\mathbf{r}}^2 \rangle = \langle \hat{\mathbf{r}}^2 \rangle - \langle \hat{\mathbf{r}} \rangle^2$ given by 
\begin{equation}
\begin{split}
    \langle \delta\hat{\mathbf{r}}^2 \rangle = \frac{1}{\mathcal{N}} \sum_{\mathbf{k}} \left| i \nabla_\mathbf{k} \chi_\mathbf{k} - \chi_\mathbf{k} \nabla_\mathbf{k} \theta_\mathbf{k}  \right|^2 - \frac{1}{\mathcal{N}^2} \left| \sum_\mathbf{k} \mathbf{\mathcal{A}}_\mathbf{k}\right|^2.
\end{split}
\end{equation}
It is straightforward to check that this quantity is gauge invariant under $\chi_\mathbf{k} \mapsto e^{i\vartheta_\mathbf{k}}\chi_\mathbf{k},$ which must be the case since energy eigenstates should have well-defined probability density. For a one-band model, $\chi_\mathbf{k} = 1;$ so the width of $\ket{\Psi_{\mathbf{a}_i}}$ is controlled by $|\sum_\mathbf{k}\nabla_\mathbf{k}\theta_\mathbf{k}|^2,$ which vanishes as $E^{-2}$ since $\theta_\mathbf{k}$ only depends on the energy dispersion in this case. Therefore, in a one-band model, the wavefunction becomes completely spatially localized in the strong-field limit, leading to Stark-Wannier localization. In a many-band model, the Berry connection obstructs this localization. In the large-$E$ limit, the energy eigenstates can retain a finite $E$-independent width. As argued in Ref. \cite{BCSH90}, this means that with the Berry connection, the matrix elements of the current operator between adjacent eigenstates need not vanish in the large-field limit, leading to a residual current in that regime.

A connection to the semiclassical description can be found by studying the dynamics of a wave packet initially localized at $\mathbf{a}_j$ centered around crystal momentum $\mathbf{k}_0$ 
\begin{equation}
\label{eq: wave packet def}
    \ket{\phi_{\mathbf{a}_j,\mathbf{k}_0}} = \frac{1}{A}\sum_{\mathbf{a}_i}\exp \left[i \mathbf{k}_0 \cdot \mathbf{a}_i - \frac{|\mathbf{a}_i-\mathbf{a}_j|^2}{2a^2} \right] \ket{W_{\mathbf{a}_i}},
\end{equation}
where $a$ is the spatial width of the wave packet, assumed larger than the lattice spacing to make the theory analytically controlled,  $A \approx \sqrt{\pi a^2/V},$ and $\ket{W_{\mathbf{a}_i}} = \frac{1}{\sqrt{\mathcal{N}}} \sum_{\mathbf{k}} e^{-i \mathbf{k} \cdot \mathbf{a}_i} \ket{\psi_\mathbf{k}}$ are band-projected exponentially-localized Wannier functions. To calculate time evolution, we need $\hat{\mathcal{U}}(t) = \sum_{\mathbf{a}_i} \ket{\Psi_{\mathbf{a}_i}}\bra{\Psi_{\mathbf{a}_i}} e^{-ie \mathbf{E}\cdot \mathbf{a}_i t/\hbar}$ which has matrix elements
\begin{equation}
\label{eq: matrix elements of U}
    \bra{W_{\mathbf{a}_i'}}\hat{\mathcal{U}}(t) \ket{W_{\mathbf{a}_i}} = \frac{1}{\mathcal{N}} \sum_{\mathbf{k}} e^{i\theta_{\mathbf{k}(t)} -i\theta_\mathbf{k}+i \mathbf{k}(t) \cdot \mathbf{a}_i'- i \mathbf{k} \cdot \mathbf{a}_i },
\end{equation}
where $\mathbf{k}(t) = \mathbf{k} - e \mathbf{E}t/\hbar$, which reproduces the acceleration theorem requiring  $\mathbf{k}\mapsto\mathbf{k} - e \mathbf{E}t/\hbar$. We calculate the time evolution of the position operator $\langle\hat{\mathbf{r}}(t) \rangle$  expanded in the Wannier basis \cite{MV97}
\begin{equation}
\label{eq: time evolution of r}
\begin{split}
    \langle\hat{\mathbf{r}}(t) \rangle &= \sum_{\mathbf{a}_i',\mathbf{a}_i''}\bra{\phi_{\mathbf{a}_j,\mathbf{k}_0}}\hat{\mathcal{U}}^\dagger(t)\ket{W_{\mathbf{a}_i'}}\times \\
    &\times \bra{W_{\mathbf{a}_i'}}\hat{\mathbf{r}}\ket{W_{\mathbf{a}_i''}}\bra{W_{\mathbf{a}_i''}}\hat{\mathcal{U}}(t)\ket{\phi_{\mathbf{a}_j,\mathbf{k}_0}}, \\
    \bra{W_{\mathbf{a}_i'}}\hat{\mathbf{r}}\ket{W_{\mathbf{a}_i''}}  &= \frac{1}{\mathcal{N}} \sum_{\mathbf{k}}\boldsymbol{\mathcal{A}}_{\mathbf{k}} e^{i \mathbf{k}\cdot \left(\mathbf{a}_i'-\mathbf{a}_i'' \right)} + \mathbf{a}_i'' \delta_{\mathbf{a}_i',\mathbf{a}_i''}.
\end{split}
\end{equation}
Then, differentiating $\partial_t \langle \hat{\mathbf{r}}(t) \rangle,$ we find 
\begin{equation}
   \hbar \frac{d\langle\hat{\mathbf{r}}(t) \rangle}{dt} = \nabla_{\mathbf{k}}\varepsilon_{\mathbf{k}(t)} - e\mathbf{E} \times  \boldsymbol{\Omega}_{\mathbf{k}(t)},
\end{equation}
which recovers exactly Eq. \eqref{eq: semiclassical dynamics}.

We now assess the consequence of Bloch and geometric oscillations on transport properties. To do this, we use the steady-state Boltzmann equation to find the occupation function $f_\mathbf{k}= \sum_{\mathbf{a}_i} \left(1-ie\mathbf{E} \cdot \mathbf{a}_i \tau/\hbar\right)^{-1}\tilde{f}^{0}_{\mathbf{a}_i} e^{i \mathbf{k}\cdot \mathbf{a}_i}$ from the equilibrium occupation $f^0_\mathbf{k} = \sum_{\mathbf{a}_i} \tilde{f}^0_{\mathbf{a}_i} e^{i \mathbf{k} \cdot \mathbf{a}_i},$  where $\tau$ is the relaxation time, which we assume is a constant \cite{P65, AM76, FDKL21}. We find two distinct regimes:
\begin{equation}
\label{eq: small E occupation}
    f_\mathbf{k} \approx f^0_\mathbf{k}+\frac{e\tau}{\hbar} \mathbf{E} \cdot \nabla_\mathbf{k} f^0_\mathbf{k}
\end{equation}
for small $\mathbf{E},$ and 
\begin{equation}
\label{eq: large E occupation}
    f_\mathbf{k} \approx \sum_{\mathbf{E} \cdot \mathbf{a}_i = 0}\tilde{f}^0_{\mathbf{a}_i} e^{i \mathbf{k} \cdot \mathbf{a}_i}+\frac{i\hbar}{e\tau} \sum_{\mathbf{E} \cdot \mathbf{a}_i \neq 0} \left(\mathbf{E} \cdot \mathbf{a}_i\right)^{-1} \tilde{f}^0_{\mathbf{a}_i}e^{i \mathbf{k} \cdot \mathbf{a}_i}. 
\end{equation}
for large $\mathbf{E}.$ The steady-state drift currents are given by $\mathbf{J}_\text{Bloch} = -\frac{e}{\hbar} \int \frac{d^2 \mathbf{k}}{(2\pi)^2}  f_\mathbf{k}  \nabla_\mathbf{k} \varepsilon_\mathbf{k}$ and $\mathbf{J}_\text{geom} = \frac{e^2}{\hbar} \int \frac{d^2 \mathbf{k}}{(2\pi)^2} \mathbf{E} \times\left(  f_\mathbf{k} \boldsymbol{\Omega}_\mathbf{k}\right).$ Assuming time-reversal symmetry in the absence of an external field throughout this work, we have $f_\mathbf{k}^0 = f_{-\mathbf{k}}^0$ and $\tilde{f}_{\mathbf{a}_i}^0 = \tilde{f}_{-\mathbf{a}_i}^0.$ Using this, we see that the first terms in Eqs. \eqref{eq: small E occupation} and \eqref{eq: large E occupation} are even under time reversal, while the second terms are odd. Since both $\nabla_\mathbf{k} \varepsilon_\mathbf{k}$ and $\boldsymbol{\Omega}_\mathbf{k}$ are odd under time reversal, only the odd components of the occupation function contribute to the current density. We  observe further that the $\mathcal{T}$-odd components of $f_\mathbf{k}$ are also odd under $\mathbf{E} \mapsto - \mathbf{E}$ in both the small and large field limits. This means that $\mathbf{J}_\text{Bloch}$ is generically odd under $\mathbf{E}$-inversion, while $\mathbf{J}_\text{geom}$ is even under $\mathbf{E}$-inversion in these two regimes. This symmetry property was also found by the authors of Ref. \cite{PC12}. In steady state, the currents are time-independent; so Bloch and geometric oscillations do not manifest explicitly; instead, their effects show up in the scaling behaviors of these currents \footnote{It is in principle possible to observe oscillatory currents if the sample is clean enough that the scattering time is at least several times the period of the oscillations. However, these transient signatures are more difficult to achieve experimentally so we focus instead on steady-state effects in this work. We leave the analysis of the time-domain signals to future work}.

In the small field limit, at lowest order, $|\mathbf{J}_\text{Bloch}|$ scales as $E = |\mathbf{E}|;$ this is the usual Drude regime  where current varies linearly with applied voltage. On the other hand, $|\mathbf{J}_\text{geom}|$ scales as $E^2$ at small fields because the occupation function contributes one factor of $E$ and the curvature-induced velocity contributes another. This is the conventional second-order nonlinear Hall effect \cite{LJG15,SF15}. If symmetry forbids the second-order signal, then the nonlinear Hall effect is generically activated at higher orders in $E$. Regardless, this small $E$ regime contains no information about the oscillatory behavior predicted by the semiclassical equations. To observe such an oscillation, we must go to the large-field regime where the scaling behavior is completely different. Here, we have
\begin{equation}
    \begin{split}
        \mathbf{J}_\text{Bloch} &\approx -\frac{i}{\tau} \sum_{\mathbf{E} \cdot \mathbf{a}_i \neq 0}\tilde{f}_{\mathbf{a}_i}^0 \int \frac{d^2 \mathbf{k}}{(2\pi)^2} \frac{\nabla_\mathbf{k}\varepsilon_\mathbf{k}}{ \mathbf{E} \cdot \mathbf{a}_i} e^{i \mathbf{k} \cdot \mathbf{a}_i}   , \\
         \mathbf{J}_\text{geom} &\approx \frac{ie}{\tau} \sum_{\mathbf{E} \cdot \mathbf{a}_i \neq 0} \tilde{f}_{\mathbf{a}_i}^0 \int \frac{d^2 \mathbf{k}}{(2\pi)^2} \frac{\hat{\mathbf{e}}_\mathbf{E}\times \boldsymbol{\Omega}_\mathbf{k}}{\hat{\mathbf{e}}_\mathbf{E} \cdot \mathbf{a}_i}  e^{i\mathbf{k} \cdot  \mathbf{a}_i},  \\
    \end{split}
\end{equation}
where $\hat{\mathbf{e}}_{\mathbf{E}} = \mathbf{E}/E.$ $\mathbf{J}_\text{Bloch}$ decays as $E^{-1}$. On the other hand, $\mathbf{J}_\text{geom}$ generically reaches a constant residual current density that is independent of $E$. If this residual current is nonzero, it means that complete Wannier-Stark localization is impossible since the current does not vanish in large $E$. Therefore, a signature of geometric quantum oscillations is \textit{vanishing} differential \textit{Hall} conductance in the infinite $E$ limit.  This is the geometric analog to negative \textit{longitudinal} differential conductance that is believed to give an unambiguous signature of the usual Bloch oscillations \cite{ET70,BCSH90, SPWM90}. When $\mathbf{E}$ crosses over from the regime of quadratically-amplified $\mathbf{J}_\text{geom}$ to the regime of constant $\mathbf{J}_\text{geom},$ this can be accompanied by a region in $E$-space where negative differential Hall conductance is observed, although this is not guaranteed in general.

\begin{figure}
    \centering
    \includegraphics[width=3in]{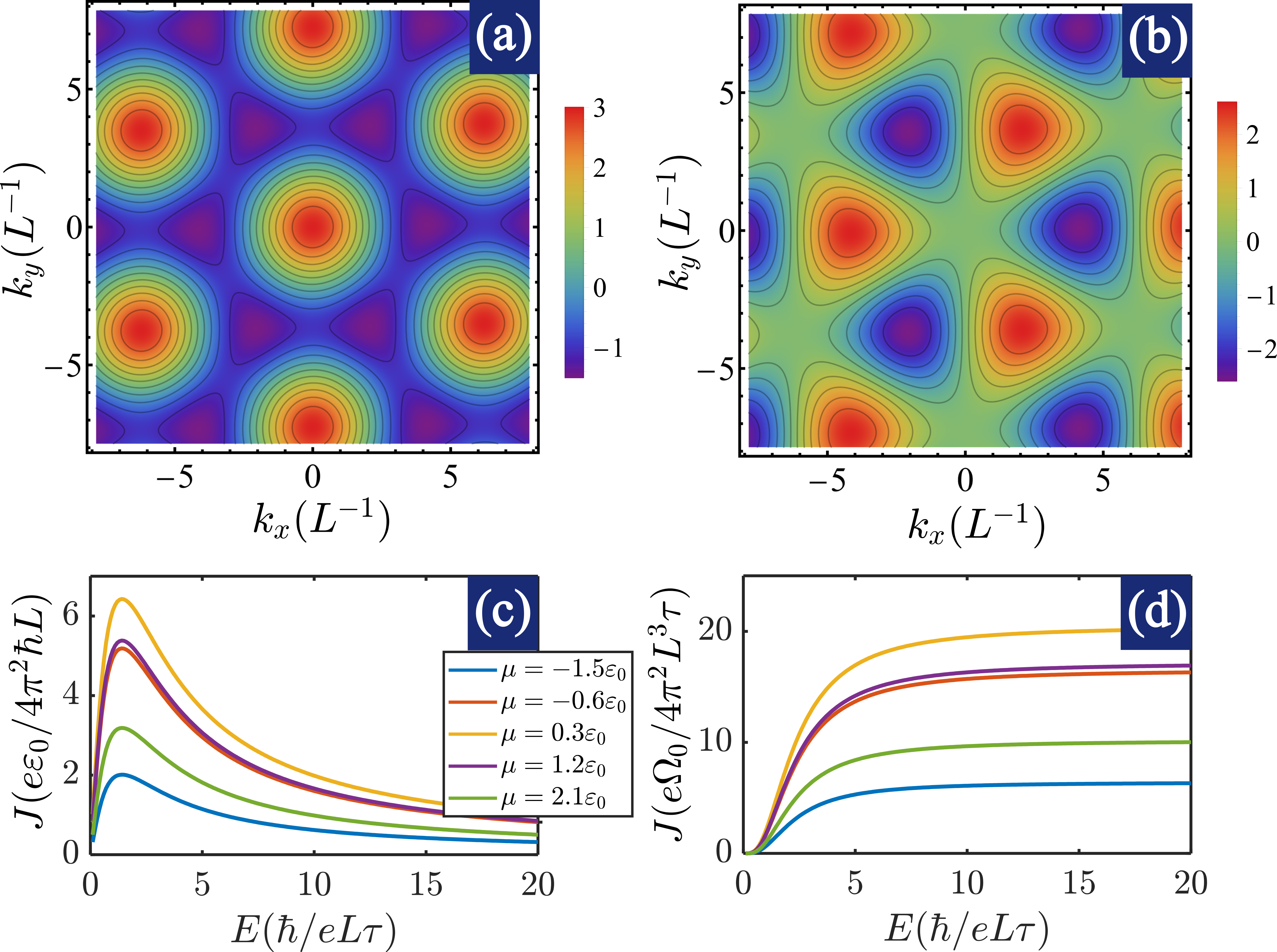}
    \caption{\textbf{Bloch and geometric oscillations in model system.} (a)-(b) Band structure and Berry curvature of model system defined in the text. This system respects $\mathcal{T}$ symmetry and $\mathcal{C}_{3z}$ rotation symmetry. The Berry curvature is even (odd) under  $\mathcal{M}_y$ ($\mathcal{M}_x$) mirror exchange  (c) $\mathbf{J}_\text{Bloch} = J \hat{\mathbf{e}}_\mathbf{x}$ as a function of applied electric field $\mathbf{E} = E\hat{\mathbf{e}}_\mathbf{x}$ for various values of the chemical potential. Onset of negative differential conductance occurs around $E \approx 1.$ (d) $\mathbf{J}_\text{geom} =  J \hat{\mathbf{e}}_\mathbf{y}$ as a function of applied electric field $\mathbf{E} = E\hat{\mathbf{e}}_\mathbf{x}$ for various values of the chemical potential. In this case, there is no negative differential conductance; instead, the current plateaus at some finite value for large $E.$ }
    \label{fig:model system}
\end{figure}

The  physics discussed  above can be illustrated using a simple model with energy dispersion $\varepsilon(\mathbf{k}) = \varepsilon_0 \sum_{i=1}^3 \cos \left( \mathbf{k} \cdot \mathbf{a}_i \right)$ and Berry curvature $\boldsymbol{\Omega}(\mathbf{k}) = \Omega_0 \hat{\mathbf{e}}_\mathbf{z} \sum_{i=1}^3 \sin \left( \mathbf{k} \cdot \mathbf{a}_i \right),$ where $\mathbf{a}_1 = L \left(-\frac{1}{2}, \frac{\sqrt{3}}{2} \right),$ $\mathbf{a}_2 = L \left(1,0 \right),$ and $\mathbf{a}_3 = - \mathbf{a}_1-\mathbf{a}_2.$ As shown in Fig. \ref{fig:model system}(a)-(b), this model respects time-reversal symmetry as well as threefold rotation symmetry; the Berry curvature is even (odd) under mirror exchange about the $x$-axis ($y$-axis). It importantly breaks inversion symmetry because otherwise in combination with time-reversal symmetry, the Berry curvature would be required to vanish. From a symmetry point-of-view, this model is equivalent to monolayer graphene with a sublattice staggered potential. If one projects to the valence or conduction band  in that system and keep only the first star of Fourier harmonics in the energy dispersion and Berry curvature, one would obtain the present toy model. 
 
\begin{figure}
    \centering
    \includegraphics[width=3in]{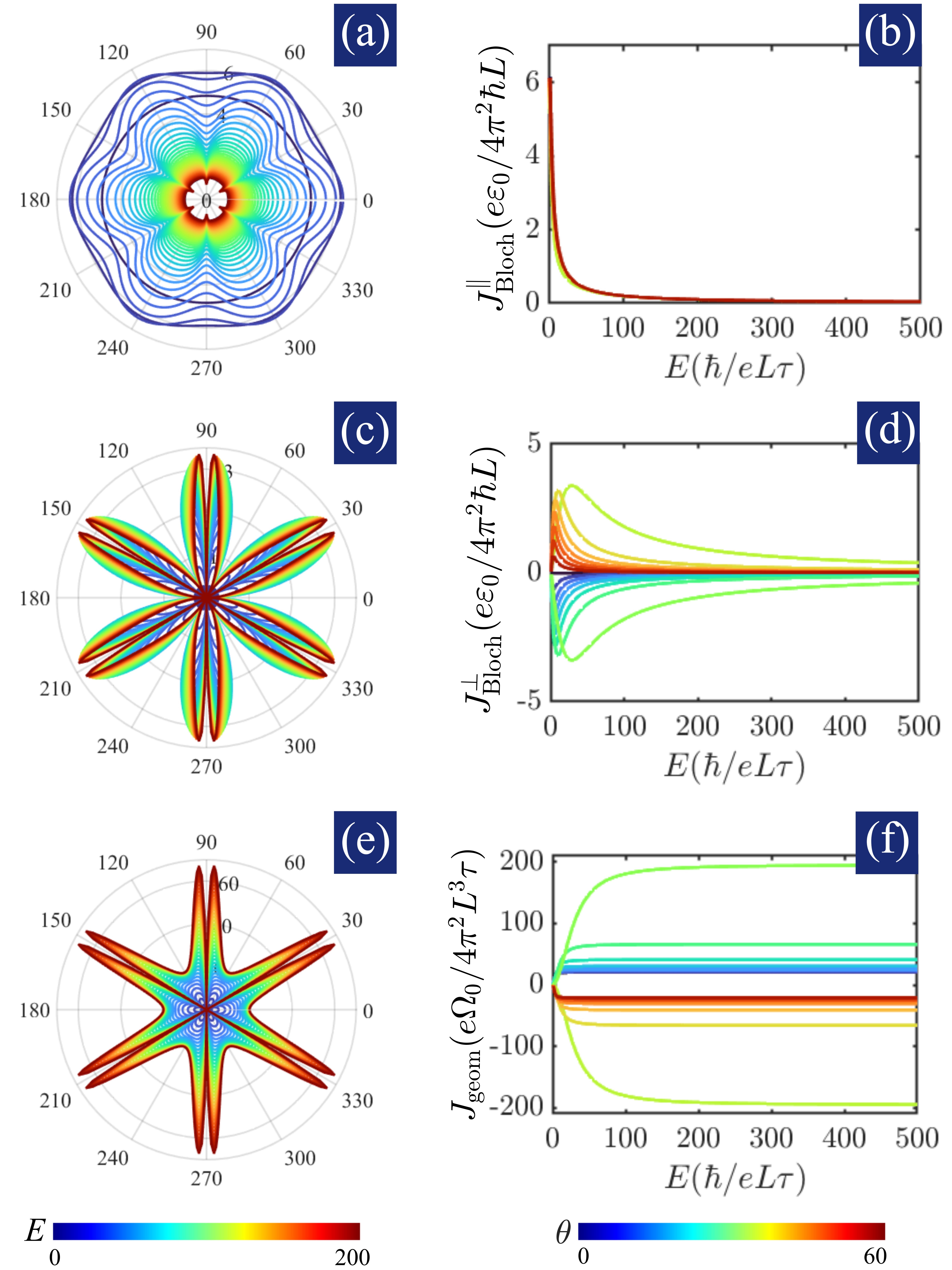}
    \caption{\textbf{Angular anisotropy of currents.} Polar plots of the magnitudes (a) $|J^\parallel_\text{Bloch}|,$ (c) $|J^\perp_\text{Bloch}|,$ and (e) $|J_\text{geom}|$ with the angle coordinate corresponding to the angle of $\mathbf{E},$ the radius coordinate corresponding to the current magnitude, and color corresponding to the magnitude of $\mathbf{E}.$ Line plots showing (b) $J^\parallel_\text{Bloch},$ (d) $J^\perp_\text{Bloch},$ and (f) $J_\text{geom}$ as functions of $E$ for different values of $0^\circ \leq \theta \leq 60^\circ$. While $J^\parallel_\text{Bloch}$ is only weakly dependent on $\theta,$ the transverse currents $J^\perp_\text{Bloch}$ and $J_\text{geom}$ feature strong angular anisotropy. However, $J^\perp_\text{Bloch}$ is even under $\mathbf{E}$-inversion, while $J_\text{geom}$ is odd under $\mathbf{E}$-inversion. }
    \label{fig: directional dependence of model system}
\end{figure}

Applying an electric field along the $x$-direction, we find $\mathbf{J}_\text{Bloch}$ points entirely along $\mathbf{E},$ with magnitude that first increases linearly with $E,$ peaks around $eEL\tau/\hbar \approx 1,$ and then monotonically decreases with further increases in $E,$ as shown in Fig. \ref{fig:model system}(c). On the other hand, $\mathbf{J}_\text{geom}$ is aligned orthogonally to $\mathbf{E}$ with magnitude that increases rapidly as a nonlinear power law at small $E,$ then  changes concavity and approaches asymptotically a constant limit at large $E,$ as shown Fig. \ref{fig:model system}(d). Next, we assess the dependence of the currents on the direction of $\mathbf{E},$ which we write as $\mathbf{E} = E \cos \theta \hat{\mathbf{e}}_\mathbf{x} + E \sin \theta \hat{\mathbf{e}}_\mathbf{y}.$  We decompose $\mathbf{J}_\text{Bloch} = J^\parallel_\text{Bloch} \hat{\mathbf{e}}_{\mathbf{E}} + J^\perp_\text{Bloch} \hat{\mathbf{e}}_\mathbf{z} \times \hat{\mathbf{e}}_{\mathbf{E}}$ and $\mathbf{J}_\text{geom} = J_\text{geom} \hat{\mathbf{e}}_\mathbf{z} \times \hat{\mathbf{e}}_{\mathbf{E}}.$ For a general $\theta,$ both $J^\parallel_\text{Bloch}$ and $J^\perp_\text{Bloch}$ are non-zero at small $E$ and decay to zero at large $E$, as shown in Fig. \ref{fig: directional dependence of model system}(a)-(d). However, the crossover from the small-$E$ to the large-$E$ regimes is only weakly dependent on $\theta$ for $J^\parallel_\text{Bloch},$ while $J^\perp_\text{Bloch}$
features strong aniostropy in this crossover behavior. $J_\text{geom}$ also features strong aniostropy; the residual current at large $E$ diverges near $30^\circ, 90^\circ, 150^\circ, 210^\circ, 270^\circ, \text{ and } 330^\circ.$ The two transverse currents are distinguishable by their behavior under $\mathbf{E}$-inversion; $J^\perp_\text{Bloch}$ is even under $\mathbf{E}$-inversion, while $J_\text{geom}$ is odd under $\mathbf{E}$-inversion \footnote{There is a potential confusion in  this terminology. $\mathbf{J}^\perp_\text{Bloch}$ is odd under $\mathbf{E}$-inversion, while $\mathbf{J}_\text{geom}$ is even under $\mathbf{E}$-inversion. However, since the basis vector $\hat{\mathbf{e}}_\mathbf{E}$ is odd under $\mathbf{E}$-inversion, the \textit{components} $J^\perp_\text{Bloch}$ and $J_\text{geom}$ obey the opposite symmetry constraints to their vectors.}.

To observe geometric oscillations, at least three requirements must be met: (1) the lattice constant must be large enough that the onset of geometric oscillations $eEL\tau/\hbar > 1$ can be achieved with reasonable field strenghs, (2) the bandwidth must be small enough compared to the band gaps that Zener tunneling can be neglected $eEL \ll \mathcal{E}_\text{gap}^2/\mathcal{E}_\text{width}$ \cite{AM76}, and (3) the bands of interest must carry significant Berry curvature. All of these requirements can be met with superlattice materials that host flat bands. For example, twisted bilayer graphene aligned on hexagonal boron nitride, which induces a $\sigma_z$ mass on the bottom graphene layer of about $30$ meV \cite{KF18, LLB21, LPZ22},  is a promising candidate. At $\theta = 1.0^\circ$ with $L \approx 140$ \AA, each valley carries a Chern band with $\mathcal{C} = \pm 1$ and $\mathcal{E}_\text{gap}^2/\mathcal{E}_\text{width} \approx 5.5$ meV. For a typical relaxation time of $1$ ps \cite{DSC08}, the onset of geometric oscillations requires $E \approx 0.5$ kV/cm, which is both experimentally feasible and is well below the Zener limit. Beyond twisted bilayer graphene, the many recently-discovered twistronic materials are promising candidates. For example, it is possible to increase the valley Chern number in graphene-based flat bands by stacking more layers, which would enhance the nonlinear Hall current \cite{ZMJS19, CCJ19, K19, LMGD19, HWKY20}. At small fields, the second-order nonlinear Hall effect has already been observed in twisted bilayer and double bilayer graphene \cite{DJPZ22,SAC22,Chakraborty_2022}. It is reasonable to speculate that at larger fields, the onset of geometric oscillations might be observable in these platforms.

In closing, we comment on several possible extensions. Of immediate relevance is augmenting the present discussion to include  Zener tunneling in which Berry phase effects are known to appear \cite{SNT20}\footnote{Presumably, the many-band formalism would also lift the restriction that the band of interest be Chern trivial as assumed in this work since in such a more general framework, only the sum of all Chern numbers needs to be zero.}.  It would also be interesting to consider the signatures of geometric oscillations in experimental probes beyond transport measurements, such as in optical setups \cite{FLSM92,L98}. Finally, Bloch oscillations have been extensively studied both experimentally and theoretically in non-solid-state media, including in cold atoms systems \cite{BPRCS96} and in photonic waveguides \cite{PPL98,MPAES99,SCWGOP03,L09}. Perhaps these platforms can also host geometric oscillations once the right ingredients have been added to proliferate the Berry curvature.

\begin{acknowledgments}
We thank Christophe De Beule for valuable conversations. This work is supported by the Department of Energy under grant DE-FG02-84ER45118. VTP acknowledges further support from the P.D. Soros Fellowship for New Americans and the National Science Foundation's Graduate Research Fellowships Program. 
\end{acknowledgments}

\bibliography{references.bib}

\newpage
\onecolumngrid
\setcounter{equation}{0}
\setcounter{figure}{0}
\renewcommand{\theequation}{S\arabic{equation}}
\renewcommand{\thefigure}{S\arabic{figure}}
\renewcommand{\bibnumfmt}[1]{[#1]}
\renewcommand{\citenumfont}[1]{#1}

\vspace{0.5in}

\begin{center}
\begin{Large}
Supplementary Material 
\end{Large}
\end{center}

\section{Quantum Dynamics in the Wannier-Stark Formalism}

In this section, we provide details on the Wannier-Stark model. Here, we will adopt a slightly different set of notations compared to those used in the main text. Whereas simplified notations are employed in the main text to maintain brevity, explicit and unambiguous, though at the risk of being verbose, notations are used here. The translation between the two sets of notations should be straightforward. We use the tight-binding framework on a lattice with $\sigma$ orbitals per unit cell and denoted by $\ket{\mathbf{a}_i,\sigma},$ where $\mathbf{a}_i$ are the translation vectors and $\boldsymbol{\tau}_\sigma$ are the basis vectors.  The Hamiltonian in real space, with the inclusion of a homogeneous static electric field $\mathbf{E}$, is 
\begin{equation}
    \hat{\mathcal{H}} = \hat{\mathcal{H}}_0 + \delta \hat{\mathcal{H}} =  \hat{\mathcal{H}}_0 +e \mathbf{E} \cdot \sum_{\mathbf{a}_i,\sigma} \left(\mathbf{a}_i + \boldsymbol{\tau}_\sigma \right) \ket{\mathbf{a}_i,\sigma} \bra{\mathbf{a}_i,\sigma},
\end{equation}
where $ \hat{\mathcal{H}}_0 $ is the hopping Hamiltonian in the absence of field. We have assumed that the atomic orbitals are so localized that the $\hat{\mathbf{r}}$ position operator is diagonal in this basis. In our notation, the elementary charge is $q = -e.$ Due to translational symmetry, we use the Fourier representation in which
\begin{equation}
\begin{split}
     \ket{\mathbf{k},\sigma } = \frac{1}{\sqrt{\mathcal{N}}}\sum_{\mathbf{a}_i} e^{i\mathbf{k} \cdot \left(\mathbf{a}_i + \boldsymbol{\tau}_\sigma\right)} \ket{\mathbf{a}_i ,\sigma}, \quad  \ket{\mathbf{a}_i, \sigma} =  \frac{1}{\sqrt{\mathcal{N}}}\sum_{\mathbf{k}} e^{-i\mathbf{k} \cdot \left( \mathbf{a}_i + \boldsymbol{\tau}_\sigma \right)} \ket{\mathbf{k} ,\sigma} , 
\end{split}
\end{equation}
where $\mathcal{N}$ is the number of unit cells. These states have unit normalization: $\bra{\mathbf{a}_i',\sigma'}\ket{\mathbf{a}_i,\sigma} = \delta_{\mathbf{a}_i,\mathbf{a}_i'} \delta_{\sigma,\sigma'}$ and $\bra{\mathbf{k}',\sigma'}\ket{\mathbf{k},\sigma} = \delta_{\mathbf{k},\mathbf{k}'} \delta_{\sigma,\sigma'}.$ We calculate the matrix elements of $\delta\hat{\mathcal{H}}$
\begin{equation}
\begin{split}
    &e \mathbf{E} \cdot \sum_{\mathbf{a}_i,\sigma''} (\mathbf{a}_i + \boldsymbol{\tau}_{\sigma''} )\bra{\mathbf{k}',\sigma'}\ket{\mathbf{a}_i,\sigma''} \bra{\mathbf{a}_i,\sigma''}\ket{\mathbf{k},\sigma}  = i\delta_{\sigma', \sigma}\delta_{\mathbf{k}', \mathbf{k}} e\mathbf{E}  \cdot\nabla_{\mathbf{k}}.\\ 
\end{split}
\end{equation}
This has a heuristic interpretation that the dipole operator $e\hat{\mathbf{r}}$ can be replaced by $ie \nabla_\mathbf{k}.$ We let the derivative operator act on the right. Technically, we must regard $\mathbf{k}$ as a continuous variable to justify taking the derivative. However, our definition of $\ket{\mathbf{k},\sigma}$ suggests that it is discrete, which is more convenient for numerical simulation. When $\mathcal{N}$ is macroscopically large, this ambiguity does not cause any  difficulties. As a check of consistency, we calculate
\begin{equation} 
\begin{split}
&\sum_{\mathbf{k},\sigma} i\ket{\mathbf{k},\sigma} e\mathbf{E} \cdot \nabla_{\mathbf{k}}\bra{\mathbf{k},\sigma}  = \frac{1}{\mathcal{N}}\sum_{\mathbf{k}, \mathbf{a}_i, \mathbf{a}_i',\sigma} e \mathbf{E} \cdot \left( \mathbf{a}_i + \boldsymbol{\tau}_\sigma \right) e^{i\mathbf{k} \cdot \left(\mathbf{a}_i'- \mathbf{a}_i  \right)} \ket{\mathbf{a}_i',\sigma} \bra{\mathbf{a}_i,\sigma} = \sum_{\mathbf{a}_i, \sigma} e \mathbf{E} \cdot \left( \mathbf{a}_i + \boldsymbol{\tau}_\sigma \right)  \ket{\mathbf{a}_i,\sigma} \bra{\mathbf{a}_i,\sigma}, 
\end{split}
\end{equation}
which is exactly the Hamiltonian in real-space. The expansion of $\hat{\mathcal{H}}_0$ in the $\mathbf{k}$ basis is standard
\begin{equation}
\begin{split}
    \mathcal{H}_0^{\sigma',\sigma}(\mathbf{k}) &= \bra{\mathbf{k}, \sigma'}\hat{\mathcal{H}}_0 \ket{\mathbf{k},\sigma} = \sum_{\mathbf{a}_i} e^{i \mathbf{k} \cdot \left(\mathbf{a}_i + \boldsymbol{\tau}_{\sigma}-   \boldsymbol{\tau}_{\sigma'}\right)} \bra{\mathbf{0}, \sigma'}\hat{\mathcal{H}}_0 \ket{\mathbf{a}_i,\sigma}.
\end{split}
\end{equation}
In all, the Hamiltonian is diagonal in $\mathbf{k}$
\begin{equation}
\label{eq: Hamiltonian in sublattice basis}
    \hat{\mathcal{H}} = \sum_{\mathbf{k},\sigma',\sigma} \ket{\mathbf{k},\sigma'} \left[ \mathcal{H}_0^{\sigma',\sigma}(\mathbf{k}) +i \delta_{\sigma',\sigma} e\mathbf{E} \cdot \nabla_{\mathbf{k}} \right] \bra{\mathbf{k},\sigma}.
\end{equation}
It is worth pointing out that $\mathcal{H}_0(\mathbf{k}+\mathbf{G}) \neq \mathcal{H}_0(\mathbf{k})$ for non-zero reciprocal lattice vectors $\mathbf{G}$ because of our choice of Fourier transform. Instead, $\mathcal{H}_0(\mathbf{k}+\mathbf{G}) = B(\mathbf{G}) \mathcal{H}_0(\mathbf{k}) B^\dagger(\mathbf{G}),$ where $B_{\sigma',\sigma}(\mathbf{G}) = \delta_{\sigma',\sigma} e^{-i \mathbf{G} \cdot \boldsymbol{\tau}_\sigma}$ is a unitary matrix. The basis states transform as $\ket{\mathbf{k}+\mathbf{G}, \sigma} = e^{i \mathbf{G} \cdot \boldsymbol{\tau}_\sigma} \ket{\mathbf{k}, \sigma} = \sum_{\sigma'} \left[B^\dagger(\mathbf{G})\right]_{\sigma,\sigma'} \ket{\mathbf{k},\sigma '}.$

Instead of the sublattice basis, let us write the Hamiltonian in the energy eigenbasis. We write the orthonormal eigenvectors of the Hamiltonian matrix $\mathcal{H}_0(\mathbf{k})$ as $\chi_n(\mathbf{k})$ with energies $\varepsilon_n(\mathbf{k})$
\begin{equation}
\label{eq: eigenbasis}
\begin{split}
    \mathcal{H}_0(\mathbf{k}) \chi_n(\mathbf{k}) = \varepsilon_n(\mathbf{k}) \chi_n(\mathbf{k}), \quad \ket{\psi_n(\mathbf{k})} = \sum_{\sigma} \chi_{n,\sigma} (\mathbf{k}) \ket{\mathbf{k},\sigma},\quad \ket{\mathbf{k},\sigma} = \sum_{n}   \ket{\psi_n(\mathbf{k})}\chi_{n,\sigma}^*.
\end{split}
\end{equation}
The eigenvectors are chosen such that $\chi_{n} (\mathbf{k}+\mathbf{G}) = B(\mathbf{G}) \chi_{n} (\mathbf{k}).$ Notice that $ \ket{\psi_n(\mathbf{k}+\mathbf{G})} = \sum_{\sigma} \chi_{n,\sigma}(\mathbf{k} + \mathbf{G})\ket{\mathbf{k}+ \mathbf{G}, \sigma} = \sum_{\sigma} e^{-i \mathbf{G} \cdot \boldsymbol{\tau}_{\sigma}}\chi_{n,\sigma}(\mathbf{k})e^{i \mathbf{G} \cdot \boldsymbol{\tau}_\sigma}\ket{\mathbf{k}, \sigma} = \ket{\psi_n(\mathbf{k})}.$ Substituting Eq. \eqref{eq: eigenbasis} into Eq. \eqref{eq: Hamiltonian in sublattice basis}, we find
\begin{equation}
    \begin{split}
        \hat{\mathcal{H}} &= \sum_{\mathbf{k},\sigma',\sigma, n,n'} \ket{\psi_{n'}(\mathbf{k})} \chi^*_{n',\sigma'}  \left[\mathcal{H}_0^{\sigma',\sigma}(\mathbf{k}) + i \delta_{\sigma',\sigma} e\mathbf{E} \cdot \nabla_{\mathbf{k}} \right]   \chi_{n,\sigma}\bra{\psi_n(\mathbf{k})}\\
        &= \sum_{\mathbf{k}, n, n'}  \ket{\psi_{n'}(\mathbf{k})} \left[\delta_{n',n}\left(\varepsilon_n(\mathbf{k}) + i e\mathbf{E} \cdot \nabla_\mathbf{k} \right)+e \mathbf{E} \cdot \boldsymbol{\mathcal{A}}_{n',n}(\mathbf{k})   \right] \bra{\psi_n(\mathbf{k})} \\
        \boldsymbol{\mathcal{A}}_{n',n}(\mathbf{k}) &= i \sum_{\sigma}\chi^*_{n',\sigma} (\mathbf{k}) \nabla_{\mathbf{k}} \chi_{n,\sigma} (\mathbf{k}),
    \end{split}
\end{equation}
where $ \boldsymbol{\mathcal{A}}_{n',n}(\mathbf{k})$ is the non-Abelian Berry connection \cite{K61,DVP94}. At this point, we must worry about differentiability of $\chi_{n}(\mathbf{k}).$ For Chern trivial bands, it is always possible to choose a differentiable gauge. This is not a major restriction since the presence of a Chern number actually obscures the oscillatory behavior induced by the Berry curvature with a constant drift velocity. As such, we will assume Chern triviality throughout, and also assume that the gauge chosen for the Bloch states is  a smooth gauge.

To make analytic progress, we project to the lowest energy band  using the projector $\hat{\mathcal{P}}_1 = \sum_{\mathbf{k}} \ket{\psi_1(\mathbf{k})}\bra{\psi_1(\mathbf{k})}$
\begin{equation}
\label{eq: projected Hamil}
    \hat{\mathcal{H}}_\text{eff} = \hat{\mathcal{P}}_1 \hat{\mathcal{H}} \hat{\mathcal{P}}_1 = \sum_{\mathbf{k}} \ket{\psi_1(\mathbf{k})} \left(\bar{\varepsilon}(\mathbf{k})  + i e \mathbf{E} \cdot \nabla_\mathbf{k}\right)\bra{\psi_1(\mathbf{k})},
\end{equation}
where $\bar{\varepsilon}(\mathbf{k}) = \varepsilon_1(\mathbf{k})+ e \mathbf{E} \cdot \boldsymbol{\mathcal{A}}_{1,1}(\mathbf{k}).$ From now on, we will drop the band index $n = 1$ and just write $\ket{\psi(\mathbf{k})} = \ket{\psi_1(\mathbf{k})}$ and $\bar{\varepsilon}(\mathbf{k}) = \varepsilon(\mathbf{k})+ e \mathbf{E} \cdot \boldsymbol{\mathcal{A}}(\mathbf{k}).$ This projection is a valid approximation if there exists a wide energy gap between this band and higher energy bands, which we also assume. Eq. \eqref{eq: projected Hamil} leads to an eigenvalue equation in $\mathbf{k}$ space:
\begin{equation}
\begin{split}
 \mathcal{E}\Psi(\mathbf{k}) &=\left[\bar{\varepsilon}(\mathbf{k})  + i e \mathbf{E} \cdot \nabla_\mathbf{k}\right] \Psi(\mathbf{k}).
 \end{split}
\end{equation}
 This Hamiltonian $\mathcal{H}_\text{eff}(\mathbf{k})$ is defined on the Brilluoin zone torus because the basis states are periodic in $\mathbf{k}:$ $\ket{\psi(\mathbf{k})} = \ket{\psi(\mathbf{k} + \mathbf{G})}.$ So we demand that $\Psi(\mathbf{k}) = \Psi(\mathbf{k} + \mathbf{G}).$ Therefore, we can write $\Psi(\mathbf{k}) =\sum_{\mathbf{a}_i} e^{i\mathbf{k} \cdot \mathbf{a}_i} \tilde{\Psi}(\mathbf{a}_i)$ and find
 \begin{equation}
 \label{eq: matrix eigenvalue}
     \sum_{\mathbf{a}_i'} \tilde{\bar{\varepsilon}}(\mathbf{a}_i- \mathbf{a}_i')   \tilde{\Psi}(\mathbf{a}_i')-e \mathbf{E} \cdot  \mathbf{a}_i  \tilde{\Psi}(\mathbf{a}_i) = \mathcal{E} \tilde{\Psi}(\mathbf{a}_i).
 \end{equation}
 This is an $\mathcal{N}\times \mathcal{N}$ matrix eigenvalue equation with $\mathcal{N}$ solutions. These $\mathcal{N}$ solutions are related to each other. Suppose $\Psi(\mathbf{k})$ is a solution with $\mathcal{E}$ as the energy; then $\Psi(\mathbf{k})e^{-i \mathbf{k} \cdot \mathbf{a}_i}$ is also a solution with energy $\mathcal{E}+e \mathbf{E} \cdot \mathbf{a}_i.$ So we can generate new solutions this way by knowing just one of the solutions. Certainly, we can get $\mathcal{N}$ solutions this way. However, they are not guaranteed to be orthogonal or independent. In the restricted setting where $\mathbf{E} \cdot \mathbf{a}_i \neq 0$ for any nonzero $\mathbf{a}_i,$ this procedure does indeed produce all the states.  In principle, we can solve Eq. \eqref{eq: matrix eigenvalue} numerically. However, it is desirable to search for an analytic solution. Let us assume that $\mathcal{E} = \tilde{\bar{\varepsilon}}(\mathbf{0})$ is a solution. Then, general solutions can be written as
\begin{equation}
\begin{split}
 \Psi_{\mathbf{a}_i}(\mathbf{k}) &= \bra{\psi (\mathbf{k})}\ket{\Psi_{\mathbf{a}_i}} = \frac{1}{\sqrt{\mathcal{N}}}\exp \left(i \theta(\mathbf{k}) -i \mathbf{k} \cdot \mathbf{a}_i \right),\\
 e\mathbf{E} \cdot \nabla_\mathbf{k}\theta(\mathbf{k}) &= \bar{\varepsilon}(\mathbf{k})-\tilde{\bar{\varepsilon}}(\mathbf{0}),\\
 \mathcal{E}_{\mathbf{a}_i} &= e \mathbf{E} \cdot \mathbf{a}_i+\tilde{\bar{\varepsilon}}(\mathbf{0}).
\end{split}
\end{equation}
The function $\theta(\mathbf{k})$ can be found explicitly by expanding  $\bar{\varepsilon}(\mathbf{k})$ in a Fourier series since it is periodic in $\mathbf{k}$
\begin{equation}
\label{eq: solution}
    \begin{split}
        \bar{\varepsilon}(\mathbf{k}) &= \sum_{\mathbf{a}_i} e^{i \mathbf{a}_i \cdot \mathbf{k}} \left[\tilde{\varepsilon}(\mathbf{a}_i) + e \mathbf{E} \cdot \tilde{\boldsymbol{\mathcal{A}}}(\mathbf{a}_i)\right],  \\
        \theta(\mathbf{k}) &=  \sum_{\mathbf{a}_i  \neq \mathbf{0}}\frac{\tilde{\bar{\varepsilon}}(\mathbf{a}_i)}{ ie\mathbf{E} \cdot \mathbf{a}_i} e^{i \mathbf{a}_i \cdot \mathbf{k}} = \sum_{\mathbf{a}_i  \neq \mathbf{0}} \left[\frac{\tilde{\varepsilon}(\mathbf{a}_i) + e \mathbf{E} \cdot \tilde{\boldsymbol{\mathcal{A}}}(\mathbf{a}_i)}{ ie\mathbf{E} \cdot \mathbf{a}_i}  \right] e^{i \mathbf{a}_i \cdot \mathbf{k}} ,
    \end{split}
\end{equation}
where we have assumed that $\mathbf{E} \cdot \mathbf{a}_i \neq 0$ for any nonzero $\mathbf{a}_i = \mathbf{0}$ to avoid dividing by zero. We find it difficult to write an analytic solution when $\mathbf{E}$ is perpendicular to a lattice vector in the form of Eq. \eqref{eq: solution}. Instead, we can look to Eq. \eqref{eq: matrix eigenvalue} to find a solution numerically. We now show that  $\theta(\mathbf{k})$ in Eq. \eqref{eq: solution} is  real. Since $\bar{\varepsilon}(\mathbf{k})$ is real, $\tilde{\bar{\varepsilon}}(\mathbf{a}_i)^* = \tilde{\bar{\varepsilon}}(-\mathbf{a}_i).$ This implies that $\left[ \frac{\tilde{\bar{\varepsilon}}(\mathbf{a}_i)}{ ie\mathbf{E} \cdot \mathbf{a}_i}\right]^* = \frac{\tilde{\bar{\varepsilon}}(-\mathbf{a}_i)}{ ie\mathbf{E} \cdot (-\mathbf{a}_i)}.$ So $\theta(\mathbf{k})$ is real as well.

We now study the localization property of the wavefunction $\ket{\Psi_{\mathbf{a}_i}}$ by calculating its spatial spread
\begin{equation}
    \bra{\Psi_{\mathbf{a}_i}} |\hat{\mathbf{r}}|^2 \ket{\Psi_{\mathbf{a}_i}} - |\bra{\Psi_{\mathbf{a}_i}} \hat{\mathbf{r}}\ket{\Psi_{\mathbf{a}_i}}|^2,
\end{equation}
where 
\begin{equation}
    \ket{\Psi_{\mathbf{a}_i}} = \frac{1}{\mathcal{N}} \sum_{\mathbf{k}, \mathbf{a}_j,\sigma } \chi_{\sigma}(\mathbf{k}) e^{i \theta(\mathbf{k})} e^{i \mathbf{k} \cdot \left( \mathbf{a}_j + \boldsymbol{\tau}_\sigma -\mathbf{a}_i\right)}  \ket{\mathbf{a}_j,\sigma }.
\end{equation}
The position operator has expectation value 
\begin{equation}
    \begin{split}
        \bra{\Psi_{\mathbf{a}_i}} \hat{\mathbf{r}} \ket{\Psi_{\mathbf{a}_i}}  = \frac{1}{\mathcal{N}} \sum_{\mathbf{k}} \mathbf{\mathcal{A}}(\mathbf{k}) + \mathbf{a}_i.
    \end{split}
\end{equation}
The position is independent of the electric field and the chosen gauge for the wavefunctions, just like Wannier functions. The first term $\frac{1}{\mathcal{N}} \sum_{\mathbf{k}} \mathbf{\mathcal{A}}(\mathbf{k})$ defines the position within a unit cell $\mathbf{a}_i$. Now, we compute the second moment of the position operator
\begin{equation}
    \begin{split}
        \bra{\Psi_{\mathbf{a}_i}} \hat{\mathbf{r}}_\lambda^2 \ket{\Psi_{\mathbf{a}_i}} &=    \frac{1}{\mathcal{N}} \sum_{\mathbf{k},\sigma} \left|    i\frac{\partial \chi_\sigma(\mathbf{k})}{\partial k_\lambda}  - \chi_\sigma(\mathbf{k}) \left( \frac{\partial \theta (\mathbf{k})}{\partial k_\lambda}-  (\mathbf{a}_i)_\lambda \right)             \right|^2 \geq 0.
    \end{split}
\end{equation}
Here, $\lambda = \lbrace x, y, z \rbrace.$ Putting all of these results together, we find
\begin{equation}
\begin{split}
        \bra{\Psi_{\mathbf{a}_i}} |\hat{\mathbf{r}}|^2 \ket{\Psi_{\mathbf{a}_i}} - |\bra{\Psi_{\mathbf{a}_i}} \hat{\mathbf{r}}\ket{\Psi_{\mathbf{a}_i}}|^2 &= \frac{1}{\mathcal{N}} \sum_{\mathbf{k}} \left(- i\nabla_\mathbf{k} \chi^\dagger(\mathbf{k}) - \chi^\dagger(\mathbf{k}) \nabla_\mathbf{k} \theta(\mathbf{k})\right)\cdot \left( i\nabla_\mathbf{k} \chi(\mathbf{k}) - \chi(\mathbf{k}) \nabla_\mathbf{k} \theta(\mathbf{k})\right)- \frac{1}{\mathcal{N}^2} \left| \sum_{\mathbf{k}} \mathbf{\mathcal{A}}(\mathbf{k})\right|^2.
\end{split}
\end{equation}
Let us consider a gauge transformation of the form 
\begin{equation}
\begin{split}
     \chi(\mathbf{k}) &\mapsto \chi(\mathbf{k}) e^{i \vartheta(\mathbf{k})},   \\
     \nabla_\mathbf{k} \chi(\mathbf{k}) &\mapsto e^{i \vartheta(\mathbf{k})} \nabla_\mathbf{k} \chi(\mathbf{k}) + i\chi(\mathbf{k}) e^{i \vartheta(\mathbf{k})} \nabla_\mathbf{k} \vartheta(\mathbf{k}),\\
     \mathbf{\mathcal{A}}(\mathbf{k}) &\mapsto  \mathbf{\mathcal{A}}(\mathbf{k}) - \nabla_\mathbf{k} \vartheta(\mathbf{k}),\\
     \theta(\mathbf{k}) &\mapsto \theta(\mathbf{k}) - \vartheta(\mathbf{k}),
\end{split}
\end{equation}
where $\vartheta(\mathbf{k})$ is periodic in $\mathbf{k}.$ Using this, we see that 
\begin{equation}
\begin{split}
    \left( i\nabla_\mathbf{k} \chi(\mathbf{k}) - \chi(\mathbf{k}) \nabla_\mathbf{k} \theta(\mathbf{k})\right) &\mapsto \left( i e^{i \vartheta(\mathbf{k})}\nabla_\mathbf{k} \chi(\mathbf{k}) - \chi(\mathbf{k}) e^{i\vartheta(\mathbf{k})} \nabla_\mathbf{k}\vartheta(\mathbf{k}) - \chi(\mathbf{k}) e^{i\vartheta(\mathbf{k})} \nabla_\mathbf{k} \theta(\mathbf{k}) +  \chi(\mathbf{k}) e^{i\vartheta(\mathbf{k})} \nabla_\mathbf{k} \vartheta(\mathbf{k})\right)     \\
    & = e^{i\vartheta(\mathbf{k})}\left( i\nabla_\mathbf{k} \chi(\mathbf{k}) - \chi(\mathbf{k}) \nabla_\mathbf{k} \theta(\mathbf{k})\right).
\end{split}
\end{equation}
Thus, the quantity $\bra{\Psi_{\mathbf{a}_i}} |\hat{\mathbf{r}}|^2 \ket{\Psi_{\mathbf{a}_i}} - |\bra{\Psi_{\mathbf{a}_i}} \hat{\mathbf{r}}\ket{\Psi_{\mathbf{a}_i}}|^2$ is gauge-independent. This makes sense since energy eigenstates must have well-defined spatial probability $|\psi(\mathbf{r})|^2.$ For a one-band model, $\chi_\mathbf{k} = 1,$ so the width is entirely controlled by $\sum_\mathbf{k}|\nabla_\mathbf{k}\theta(\mathbf{k})|^2,$ which is proportional to $E^{-2}$ due to the absence of $\mathbf{\mathcal{A}}(\mathbf{k}).$ Therefore, in the strong-coupling limit, the wavefunction $\ket{\Psi_{\mathbf{a}_i}}$ becomes completely localized to a unit cell at $\mathbf{a}_i.$ In a many-band model with Berry connection, the width \textit{cannot} be eliminated by increasing the strength of the electric field.

To make connection with the semiclassical limit, we now study wave packet evolution. The time evolution operator is $(t_0 = 0)$
\begin{equation}
    \hat{\mathcal{U}}(t) = e^{-i \tilde{\bar{\varepsilon}}(\mathbf{0})t/\hbar} \sum_{\mathbf{a}_i} \ket{\Psi_{\mathbf{a}_i}}\bra{\Psi_{\mathbf{a}_i}} \exp \left(-\frac{i}{\hbar} e\mathbf{E} \cdot \mathbf{a}_i t\right).
\end{equation}
The matrix elements are 
\begin{equation}
\begin{split}
    \mathcal{U}_{\mathbf{k}',\mathbf{k}}(t) &= \bra{\psi(\mathbf{k}')} \hat{\mathcal{U}}(t) \ket{\psi(\mathbf{k})}=   \frac{e^{-i \tilde{\bar{\varepsilon}}(\mathbf{0})t/\hbar}}{\mathcal{N}}\sum_{\mathbf{a}_i}  \exp \left(i \theta(\mathbf{k}')- i \theta(\mathbf{k}) -i \left(\mathbf{k}'  - \mathbf{k}  +\frac{e\mathbf{E}  t}{\hbar}  \right)\cdot \mathbf{a}_i\right) \\
    &= e^{-i \tilde{\bar{\varepsilon}}(\mathbf{0})t/\hbar}e^{i \theta(\mathbf{k}') - i \theta(\mathbf{k})} \delta_{\mathbf{k}', \mathbf{k} - e\mathbf{E}t/\hbar},    
\end{split}
\end{equation}
which reproduce the well-known acceleration theorem. Now, let us move to the band-projected Wannier basis defined as 
\begin{equation}
    \ket{W_{\mathbf{a}_i}} = \frac{1}{\sqrt{\mathcal{N}}} \sum_{\mathbf{k}} e^{-i \mathbf{k} \cdot \mathbf{a}_i} \ket{\psi(\mathbf{k})}.
\end{equation}
We assume that $\ket{W_{\mathbf{a}_i}}$ is exponentially localized about $\mathbf{a}_i$ since a smooth gauge for $\ket{\psi(\mathbf{k})}$ has been chosen. In this Wannier basis, we have
\begin{equation}
    \begin{split}
        \mathcal{U}_{\mathbf{a}_i',\mathbf{a}_i}(t) &= \bra{W_{\mathbf{a}_i'}}\hat{\mathcal{U}}(t)\ket{W_{\mathbf{a}_i}} = \frac{1}{\mathcal{N}} \sum_{\mathbf{k}',\mathbf{k}} \mathcal{U}_{\mathbf{k}',\mathbf{k}}(t) e^{i \mathbf{k}' \cdot \mathbf{a}_i' - i \mathbf{k} \cdot \mathbf{a}_i} = \frac{e^{-i \tilde{\bar{\varepsilon}}(\mathbf{0})t/\hbar}}{\mathcal{N}} \sum_{\mathbf{k}'=\mathbf{k}-e\mathbf{E}t/\hbar} e^{i \theta(\mathbf{k}') -i\theta(\mathbf{k})+i \mathbf{k}' \cdot \mathbf{a}_i'- i \mathbf{k} \cdot \mathbf{a}_i }.
    \end{split}
\end{equation}
We are interested in the evolution of a wave packet initially localized at $\mathbf{a}_j$ with crystal momentum $\mathbf{k}_0$
\begin{equation}
    \ket{\phi_{\mathbf{a}_j,\mathbf{k}_0}} = \frac{1}{A}\sum_{\mathbf{a}_i}\exp \left[i \mathbf{k}_0 \cdot \mathbf{a}_i - \frac{|\mathbf{a}_i-\mathbf{a}_j|^2}{2a^2} \right] \ket{W_{\mathbf{a}_i}}.
\end{equation}
Here, $A$ is a dimensionless normalization constant. It is approximately $A \approx \sqrt{\pi a^2/V},$ where $V$ is the volume of a unit cell. The evolution of this state is 
\begin{equation}
    \begin{split}
        &\phi_{\mathbf{a}_j, \mathbf{k}_0}(\mathbf{a}_i',t) = \bra{W_{\mathbf{a}_i'}}\hat{\mathcal{U}}(t)\ket{\phi_{\mathbf{a}_j,\mathbf{k}_0}} = \frac{e^{-i \tilde{\bar{\varepsilon}}(\mathbf{0})t/\hbar}}{A\mathcal{N}} \sum_{\mathbf{a}_i,\mathbf{k}'=\mathbf{k}-e\mathbf{E}t/\hbar} e^{i \theta(\mathbf{k}') -i\theta(\mathbf{k})+i \mathbf{k}' \cdot \mathbf{a}_i'+i (\mathbf{k}_0 -\mathbf{k})\cdot \mathbf{a}_i - \frac{|\mathbf{a}_i-\mathbf{a}_j|^2}{2a^2}} \\
        &\approx \frac{2\pi a^2}{AV\mathcal{N}} e^{-i \tilde{\bar{\varepsilon}}(\mathbf{0})t/\hbar} \sum_{\mathbf{k}'=\mathbf{k}-e\mathbf{E}t/\hbar} \exp \left[i \left[ \theta(\mathbf{k}') -\theta(\mathbf{k})+ \mathbf{k}' \cdot \mathbf{a}_i'+ (\mathbf{k}_0 -\mathbf{k})\cdot \mathbf{a}_j\right] - \frac{a^2|\mathbf{k}-\mathbf{k}_0|^2}{2} \right],
    \end{split}
\end{equation}
where we have replaced  summation with integration $\sum_{\mathbf{a}_i} \mapsto \frac{1}{V} \int d^2 \mathbf{r}$ in order to compute the Gaussian sum approximately. This approximation introduces an ambiguity in the domain of $\mathbf{k}$ integration since the Gaussian prefactor is not periodic in $\mathbf{k};$ so different chosen Brillouin zones will give different  results. Therefore, we must choose the Brillouin zone closest to $\mathbf{k}_0$ to minimize error. As such, we rewrite 
\begin{equation}
    \phi_{\mathbf{a}_j, \mathbf{k}_0}(\mathbf{a}_i',t)= \frac{2\pi a^2}{AV\mathcal{N}} e^{-i \tilde{\bar{\varepsilon}}(\mathbf{0})t/\hbar} \sum_{\mathbf{k}'=\mathbf{k}-e\mathbf{E}t/\hbar} \exp \left[i \left[\theta\left(\mathbf{k}'+\mathbf{k}_0\right) -\theta\left(\mathbf{k}+\mathbf{k}_0\right)+ \left(\mathbf{k}'+\mathbf{k}_0\right) \cdot \mathbf{a}_i' -\mathbf{k}\cdot \mathbf{a}_j\right] - \frac{a^2|\mathbf{k}|^2}{2} \right],
\end{equation}
where $\mathbf{k}$ is now in the first Brillouin zone. We assume that $a$ is larger than the lattice constant so $2\pi a^{-1}$ is smaller than the size of the Brillouin zone. As a check of consistency, we calculate
\begin{equation}
\begin{split}
    \phi_{\mathbf{a}_j, \mathbf{k}_0}(\mathbf{a}_i,t = 0)&\approx \frac{2\pi a^2}{AV\mathcal{N}} \sum_{\mathbf{k}} \exp \left[i \left[ \left(\mathbf{k}+\mathbf{k}_0\right) \cdot \mathbf{a}_i -\mathbf{k}\cdot \mathbf{a}_j\right] - \frac{a^2|\mathbf{k}|^2}{2} \right]\\
    &\approx \frac{ a^2}{2\pi A} \int d^2\mathbf{k} \exp \left[i \left[ \left(\mathbf{k}+\mathbf{k}_0\right) \cdot \mathbf{a}_i -\mathbf{k}\cdot \mathbf{a}_j\right] - \frac{a^2|\mathbf{k}|^2}{2} \right] = \frac{1}{A} \exp \left[i \mathbf{k}_0\cdot \mathbf{a}_i - \frac{|\mathbf{a}_i-\mathbf{a}_j|^2}{2a^2} \right]\\
\end{split}
\end{equation}
which recovers the original Gaussian state. Next, we study the dependence of the position on time. We calculate
\begin{equation}
    \begin{split}
        \bra{\phi_{\mathbf{a}_j,\mathbf{k}_0}}\hat{\mathcal{U}}^\dagger(t)\hat{\mathbf{r}}\hat{\mathcal{U}}(t)\ket{\phi_{\mathbf{a}_j,\mathbf{k}_0}} &= \sum_{\mathbf{a}_i',\mathbf{a}_i''}\bra{\phi_{\mathbf{a}_j,\mathbf{k}_0}}\hat{\mathcal{U}}^\dagger(t)\ket{W_{\mathbf{a}_i'}}\bra{W_{\mathbf{a}_i'}}\hat{\mathbf{r}}\ket{W_{\mathbf{a}_i''}}\bra{W_{\mathbf{a}_i''}}\hat{\mathcal{U}}(t)\ket{\phi_{\mathbf{a}_j,\mathbf{k}_0}}.
    \end{split}
\end{equation}
To proceed, we need the matrix elements of the position operator in the Wannier basis. It is tempting to suppose that $\hat{\mathbf{r}}$ is diagonal in this basis; this is not necessarily true for Wannier functions that have some weight in neighboring unit cells. For us, it is important to consider these contributions to obtain gauge invariance. We have \cite{MV97}
\begin{equation}
    \begin{split}
        \bra{W_{\mathbf{a}_i'}}\hat{\mathbf{r}}\ket{W_{\mathbf{a}_i''}} &= \frac{1}{\mathcal{N}} \sum_{\mathbf{k}',\mathbf{k}''} e^{i \mathbf{k}' \cdot \mathbf{a}_i' - i \mathbf{k}''\cdot \mathbf{a}_i''} \bra{\psi(\mathbf{k}')} \hat{\mathbf{r}} \ket{\psi(\mathbf{k}'')} = \frac{1}{\mathcal{N}} \sum_{\mathbf{k}}\boldsymbol{\mathcal{A}}(\mathbf{k}) e^{i \mathbf{k}\cdot \left(\mathbf{a}_i'-\mathbf{a}_i'' \right)} + \mathbf{a}_i'' \delta_{\mathbf{a}_i',\mathbf{a}_i''}.
     \end{split}
\end{equation}
To show this, we use $\bra{\psi(\mathbf{k}')} \hat{\mathbf{r}} \ket{\psi(\mathbf{k}'')} = \sum_{\sigma',\sigma''}\chi^*_{\sigma'}(\mathbf{k}')\chi_{\sigma''} (\mathbf{k}'') \bra{\mathbf{k}', \sigma'}   \hat{\mathbf{r}}  \ket{\mathbf{k}'', \sigma''} = -i \sum_{\sigma'} \chi^*_{\sigma'}(\mathbf{k}')\chi_{\sigma'}(\mathbf{k}'') \nabla_{\mathbf{k}''} \left(\delta_{\mathbf{k}',\mathbf{k''}}\right)$ with integration by parts. Using this, we obtain
\begin{equation}
    \begin{split}
        \bra{\phi_{\mathbf{a}_j,\mathbf{k}_0}}\hat{\mathcal{U}}^\dagger(t)\hat{\mathbf{r}}\hat{\mathcal{U}}(t)\ket{\phi_{\mathbf{a}_j,\mathbf{k}_0}} &=  \left[\frac{ a^2}{2\pi A} \right]^2\sum_{\mathbf{a}_i', \mathbf{a}_i''} \int d^2 \mathbf{q} \exp \left[-i \left[ \theta\left(\mathbf{q}'+\mathbf{k}_0\right) -\theta\left(\mathbf{q}+\mathbf{k}_0\right)+ \left(\mathbf{q}'+\mathbf{k}_0\right) \cdot \mathbf{a}_i' -\mathbf{q}\cdot \mathbf{a}_j\right] - \frac{a^2|\mathbf{q}|^2}{2} \right] \times \\
        & \times \left[ \frac{V}{(2\pi)^2} \int d^2\mathbf{k}\boldsymbol{\mathcal{A}}(\mathbf{k}) e^{i \mathbf{k}\cdot \left(\mathbf{a}_i'-\mathbf{a}_i'' \right)} + \mathbf{a}_i'' \delta_{\mathbf{a}_i',\mathbf{a}_i''}\right] \times  \\
        & \times \int d^2 \mathbf{p} \exp \left[i \left[ \theta\left(\mathbf{p}'+\mathbf{k}_0\right) -\theta\left(\mathbf{p}+\mathbf{k}_0\right)+ \left(\mathbf{p}'+\mathbf{k}_0\right) \cdot \mathbf{a}_i'' -\mathbf{p}\cdot \mathbf{a}_j\right] - \frac{a^2|\mathbf{p}|^2}{2} \right].
    \end{split}
\end{equation}
The terms diagonal in $\mathbf{a}_i'$ are 
\begin{equation}
\begin{split}
        \sum_{\mathbf{a}_i'}  \mathbf{a}_i'\phi_{\mathbf{a}_j, \mathbf{k}_0}(\mathbf{a}_i',t)^* \phi_{\mathbf{a}_j, \mathbf{k}_0}(\mathbf{a}_i',t) &\approx  \frac{(2\pi)^2}{iV}\left[\frac{ a^2}{2\pi A}  \right]^2  \int d^2 \mathbf{q} \exp \left[-i \left[ \theta\left(\mathbf{q}'+\mathbf{k}_0\right) -\theta\left(\mathbf{q}+\mathbf{k}_0\right)-\mathbf{q}\cdot \mathbf{a}_j\right] - \frac{a^2|\mathbf{q}|^2}{2} \right] \\
        & \times \int d^2 \mathbf{k} \exp \left[i \left[\theta\left(\mathbf{k}'+\mathbf{k}_0\right) -\theta\left(\mathbf{k}+\mathbf{k}_0\right) -\mathbf{k}\cdot \mathbf{a}_j\right] - \frac{a^2|\mathbf{k}|^2}{2} \right] \nabla_\mathbf{k} \delta^{2} \left( \mathbf{k}- \mathbf{q} \right). \\
\end{split}
\end{equation}
To simplify, we integrate by parts. The boundary term is neglected under the assumption that it is exponentially suppressed by the Gaussian tail. We find
\begin{equation}
    \begin{split}
        \sum_{\mathbf{a}_i'}  \mathbf{a}_i'\phi_{\mathbf{a}_j, \mathbf{k}_0}(\mathbf{a}_i',t)^* \phi_{\mathbf{a}_j, \mathbf{k}_0}(\mathbf{a}_i',t) &\approx \frac{i(2\pi)^2}{V}\left[\frac{ a^2}{2\pi A}  \right]^2 \int d^2 \mathbf{k} \exp \left(-a^2|\mathbf{k}|^2 \right) \nabla_\mathbf{k} \left[i \left[\theta\left(\mathbf{k}'+\mathbf{k}_0\right) -\theta\left(\mathbf{k}+\mathbf{k}_0\right) -\mathbf{k}\cdot \mathbf{a}_j\right] - \frac{a^2|\mathbf{k}|^2}{2}\right].
    \end{split}
\end{equation}
The last term in the integrand must vanish upon integration because it is odd under $\mathbf{k} \mapsto - \mathbf{k},$ which is good since that term is imaginary while the expectation value is supposed to be real. So we have
\begin{equation}
     \sum_{\mathbf{a}_i'}  \mathbf{a}_i'\phi_{\mathbf{a}_j, \mathbf{k}_0}(\mathbf{a}_i',t)^* \phi_{\mathbf{a}_j, \mathbf{k}_0}(\mathbf{a}_i',t) \approx -\frac{(2\pi)^2}{V}\left[\frac{ a^2}{2\pi A}  \right]^2 \int d^2 \mathbf{k} \exp \left(-a^2|\mathbf{k}|^2 \right) \nabla_\mathbf{k}  \left[ \theta\left(\mathbf{k}-e\mathbf{E}t/\hbar+\mathbf{k}_0\right) -\theta\left(\mathbf{k}+\mathbf{k}_0\right) -\mathbf{k}\cdot \mathbf{a}_j\right] .
\end{equation}
Differentiating with respect to $t,$ we find
\begin{equation}
\begin{split}
         &\frac{(2\pi)^2}{\hbar V }\left[\frac{ a^2}{2\pi A}  \right]^2 \int d^2 \mathbf{k} \exp \left(-a^2|\mathbf{k}-\mathbf{k}_0|^2 \right) \nabla_\mathbf{k}  \left[ e \mathbf{E} \cdot \nabla_{\mathbf{k}}\theta\left(\mathbf{k}(t)\right) \right]  \\
        &= \frac{(2\pi)^2}{\hbar V }\left[\frac{ a^2}{2\pi A}  \right]^2 \int d^2 \mathbf{k} \exp \left(-a^2|\mathbf{k}-\mathbf{k}_0|^2 \right)  \left[ \nabla_\mathbf{k} \varepsilon\left(\mathbf{k}(t)\right)- e \mathbf{E} \times \boldsymbol{\Omega}\left(\mathbf{k}(t)\right) +e \left(\mathbf{E} \cdot \nabla_\mathbf{k} \right)\boldsymbol{\mathcal{A}}(\mathbf{k}(t))  \right] .\\
\end{split}
\end{equation}
We notice the last term proportional to $(\mathbf{E}\cdot \nabla_\mathbf{k}) \boldsymbol{\mathcal{A}} $ is \textit{not} gauge invariant. The differentiation notation means $\nabla_\mathbf{k} f(\mathbf{k}(t)) = \nabla_{\mathbf{k}(t)} f(\mathbf{k}(t)) = \nabla_\mathbf{k} f(\mathbf{k}) |_{\mathbf{k} = \mathbf{k}(t)}$ since $\mathbf{k}(t) = \mathbf{k}-e\mathbf{E}t/\hbar.$ The following non-diagonal elements of the position operator come to the rescue:
\begin{equation}
    \begin{split}
         & \frac{ Va^4}{(2\pi)^4 A^2} \sum_{\mathbf{a}_i',  \mathbf{a}_i''} \int d^2 \mathbf{q} \exp \left[-i \left[ \theta\left(\mathbf{q}'+\mathbf{k}_0\right) -\theta\left(\mathbf{q}+\mathbf{k}_0\right)+ \left(\mathbf{q}'+\mathbf{k}_0\right) \cdot \mathbf{a}_i' -\mathbf{q}\cdot \mathbf{a}_j\right] - \frac{a^2|\mathbf{q}|^2}{2} \right]   \left[  \int d^2\mathbf{k}\boldsymbol{\mathcal{A}}(\mathbf{k}) e^{i \mathbf{k}\cdot \left(\mathbf{a}_i'-\mathbf{a}_i'' \right)}\right] \times  \\
        & \times \int d^2 \mathbf{p} \exp \left[i \left[ \theta\left(\mathbf{p}'+\mathbf{k}_0\right) -\theta\left(\mathbf{p}+\mathbf{k}_0\right)+ \left(\mathbf{p}'+\mathbf{k}_0\right) \cdot \mathbf{a}_i'' -\mathbf{p}\cdot \mathbf{a}_j\right] - \frac{a^2|\mathbf{p}|^2}{2} \right] \\
        &= \frac{ a^4}{VA^2}  \int d^2 \mathbf{k} \exp \left[ - a^2|\mathbf{k}+e\mathbf{E}t/\hbar-\mathbf{k}_0|^2 \right]   \boldsymbol{\mathcal{A}}(\mathbf{k}) = \frac{ a^4}{VA^2}   \int d^2 \mathbf{k} \exp \left[ - a^2|\mathbf{k}-\mathbf{k}_0|^2 \right]   \boldsymbol{\mathcal{A}}(\mathbf{k}-e\mathbf{E}t/\hbar).
    \end{split}
\end{equation}
Differentiating this with time, we obtain
\begin{equation}
\begin{split}
    \frac{ a^4}{VA^2}   \int d^2 \mathbf{k} \exp \left[ - a^2|\mathbf{k}-\mathbf{k}_0|^2 \right]   \frac{d\boldsymbol{\mathcal{A}}(\mathbf{k}-e\mathbf{E}t/\hbar)}{dt}   = - \frac{ a^4}{\hbar VA^2}   \int d^2 \mathbf{k} \exp \left[ - a^2|\mathbf{k}-\mathbf{k}_0|^2 \right]   e \left( \mathbf{E} \cdot \mathbf{\nabla}_\mathbf{k} \boldsymbol{\mathcal{A}}(\mathbf{k}(t)) \right),
\end{split}
\end{equation}
which is exactly the gauge-dependent term we found before with the opposite sign.  In total, we obtain 
\begin{equation}
    \frac{d}{dt} \bra{\phi_{\mathbf{a}_j,\mathbf{k}_0}}\hat{\mathcal{U}}^\dagger(t)\hat{\mathbf{r}}\hat{\mathcal{U}}(t)\ket{\phi_{\mathbf{a}_j,\mathbf{k}_0}} = \frac{a^2}{\pi\hbar }\int d^2 \mathbf{k} \exp \left(-a^2|\mathbf{k}-\mathbf{k}_0|^2 \right)  \left[ \nabla_\mathbf{k} \varepsilon\left(\mathbf{k}(t)\right)- e \mathbf{E} \times \boldsymbol{\Omega}\left(\mathbf{k}(t)\right) \right].
\end{equation}
We can simplify slightly by expanding the integrand  around $\mathbf{k} = \mathbf{k}_0$ and keeping only the constant term to get exactly 
\begin{equation}
    \hbar\frac{d}{dt} \bra{\phi_{\mathbf{a}_j,\mathbf{k}_0}}\hat{\mathcal{U}}^\dagger(t)\hat{\mathbf{r}}\hat{\mathcal{U}}(t)\ket{\phi_{\mathbf{a}_j,\mathbf{k}_0}} =  \nabla_\mathbf{k} \varepsilon\left(\mathbf{k}_0-e\mathbf{E}t/\hbar\right)- e \mathbf{E} \times \boldsymbol{\Omega}\left(\mathbf{k}_0-e\mathbf{E}t/\hbar\right).
\end{equation}
This shows that in a fully quantum-mechanical derivation, we are able to get the contribution of the velocity from the Berry curvature.

\section{Transport Properties}

Having demonstrated the validity of the semiclassical equations in the strong-coupling limit, we now develop the corresponding transport theory based on these equations. It is quite difficult to measure the motion of an individual wave packet. Instead, typical experiments measure current densities, conductivities, and resistivities.  All of these quantities rely on the occupation function $ f(\mathbf{k},\mathbf{r},t)$,  which changes due to the presence of an electric field. The Boltzmann equation governs this change: 
\begin{equation}
    \frac{\partial f(\mathbf{k},\mathbf{r},t)}{\partial t} - \frac{e}{\hbar} \mathbf{E} \cdot  \nabla_{\mathbf{k}} f(\mathbf{k},\mathbf{r},t) + \mathbf{v}_\mathbf{k} \cdot \nabla_\mathbf{r}  f(\mathbf{k},\mathbf{r},t) = \frac{f^0(\mathbf{k}) -  f(\mathbf{k},\mathbf{r},t)}{\tau}
\end{equation}
We are interested in the spatially-uniform steady-state case, in which this simplifies to 
\begin{equation}
\label{eq: steady state equation}
    - \frac{e\tau}{\hbar} \mathbf{E} \cdot  \nabla_{\mathbf{k}} f(\mathbf{k})+ f(\mathbf{k}) =f^0(\mathbf{k}), 
\end{equation}
where $f^0(\mathbf{k})$ is the equilibrium occupation function. We assume that $f(\mathbf{k})$ is periodic in $\mathbf{k}$ to find
\begin{equation}
   \left(1 - \frac{ie\tau \mathbf{E} \cdot \mathbf{a}_i}{\hbar }  \right)\tilde{f}_{\mathbf{a}_i}  = \tilde{f}^0_{\mathbf{a}_i}.
\end{equation}
So the occupation function is 
\begin{equation}
    f(\mathbf{k}) = \sum_{\mathbf{a}_i} \frac{\tilde{f}^0_{\mathbf{a}_i}e^{i \mathbf{k} \cdot \mathbf{a}_i}}{1-ie\tau \mathbf{E}\cdot \mathbf{a}_i/ \hbar} = f^0_\mathbf{0} +\sum_{\mathbf{a}_i \neq \mathbf{0}} \frac{\tilde{f}^0_{\mathbf{a}_i}e^{i \mathbf{k} \cdot \mathbf{a}_i}}{1-ie\tau \mathbf{E}\cdot \mathbf{a}_i/ \hbar}.
\end{equation}
As a check of consistency, if an entire band is filled, then $\tilde{f}^0_{\mathbf{a}_i} = 1$ for $\mathbf{a}_i = \mathbf{0}$ and $\tilde{f}^0_{\mathbf{a}_i} = 0$ for $\mathbf{a}_i \neq \mathbf{0}.$ In this case, $f(\mathbf{k}) = f^0(\mathbf{k}),$ which makes sense because an entirely filled band is inert. When $\mathbf{E} = \mathbf{0},$ then we also have $f(\mathbf{k}) = f^0(\mathbf{k}),$ which also makes sense because without an electric field, there is no perturbation to change the occupation of states. For small $x = e\tau \mathbf{E}\cdot \mathbf{a}_i/\hbar,$ we have
\begin{equation}
    \frac{1}{1-ix} \approx 1+ix-x^2-ix^3+x^4+...
\end{equation}
This limit is reached for small fields and $f^0(\mathbf{k})$ is smooth enough that only the first few harmonics are significant. Using this expansion, we get
\begin{equation}
    f(\mathbf{k}) = f^0(\mathbf{k}) + \frac{e\tau}{\hbar}\mathbf{E} \cdot \nabla_\mathbf{k}f^{0}(\mathbf{k}) + \frac{e^2\tau^2}{\hbar^2} E_iE_j \partial_i\partial_j f^0(\mathbf{k}) + ... = \sum_{i=0}^\infty \left(\frac{e\tau}{\hbar} \right)^i \prod_{j=0}^i E_{x_j}\partial_{x_j} f^0(\mathbf{k}),
\end{equation}
where repeated spatial indices are summed. This is the limit of the conventional zero-frequency nonlinear Hall effect much discussed in the literature. Now, in the large field limit, we have a completely different behavior
\begin{equation}
    f(\mathbf{k}) \approx \sum_{\mathbf{E} \cdot \mathbf{a}_i = 0}\tilde{f}^0_{\mathbf{a}_i}e^{i \mathbf{k}\cdot \mathbf{a}_i} + \frac{i\hbar}{e\tau} \sum_{\mathbf{E} \cdot \mathbf{a}_i \neq \mathbf{0}} \frac{\tilde{f}_{\mathbf{a}_i}^0 e^{i\mathbf{k} \cdot \mathbf{a}_i}}{ \mathbf{E} \cdot \mathbf{a}_i}.
\end{equation}
The current density is given by 
\begin{equation}
\begin{split}
    \mathbf{J} &=\mathbf{J}_\text{Bloch} + \mathbf{J}_\text{geom},    \\
    \mathbf{J}_\text{Bloch} &=-\frac{e}{\hbar}\int \frac{d^2\mathbf{k}}{(2\pi)^2} \nabla_\mathbf{k} \varepsilon(\mathbf{k})  f(\mathbf{k}),  \\
    \mathbf{J}_\text{geom} &= \frac{e^2}{\hbar}\int \frac{d^2\mathbf{k}}{(2\pi)^2}  \mathbf{E} \times \boldsymbol{\Omega}(\mathbf{k})  f(\mathbf{k}).
\end{split}
\end{equation}

\section{Numerical Derivatives}

In an infinite system, it is convenient to solve Eq. \eqref{eq: steady state equation} by differentiating a Fourier series. However, in a finite system, this procedure breaks the periodicity of $f(\mathbf{k})$ in $\mathbf{k}$-space. To maintain periodicity, we must solve a finite difference equation instead but still using Fourier series. First, we do a coordinate transformation
\begin{equation}
\begin{split}
\begin{pmatrix}
        \hat{g}_1 \\
        \hat{g}_2
\end{pmatrix} =  \begin{pmatrix}
        0 & 1 \\
        - \frac{\sqrt{3}}{2} & - \frac{1}{2}
\end{pmatrix}\begin{pmatrix}
        \hat{x} \\
        \hat{y}
\end{pmatrix}.
 \end{split}
\end{equation}
Any vector can be written as $k_1 \hat{g}_1 + k_2 \hat{g}_2 = -\frac{\sqrt{3}}{2} k_2 \hat{x} + \left(k_1- \frac{1}{2}k_2 \right) \hat{y}   = k_x \hat{x}+ k_y \hat{y}.$ Consequently, the transformation of the components is 
\begin{equation}
\begin{split}
\begin{pmatrix}
        k_x \\
        k_y
\end{pmatrix} =  \begin{pmatrix}
        0 & - \frac{\sqrt{3}}{2} \\
        1 & - \frac{1}{2}
\end{pmatrix}\begin{pmatrix}
        k_1 \\
        k_2
\end{pmatrix} \mapsto \begin{pmatrix}
        k_1 \\
        k_2
\end{pmatrix} =  \begin{pmatrix}
        -\frac{1}{\sqrt{3}}& 1 \\
        -\frac{2}{\sqrt{3}} & 0
\end{pmatrix}\begin{pmatrix}
        k_x \\
        k_y
\end{pmatrix}.
 \end{split}
\end{equation}
Using this, we can rewrite
\begin{equation}
\begin{split}
        -\frac{e\tau E_x}{\hbar} \frac{\partial f}{\partial k_x}-\frac{e\tau E_y}{\hbar}  \frac{\partial f}{\partial k_y} + f =  -\frac{e\tau E_x}{\hbar} \left(-\frac{1}{\sqrt{3}}\frac{\partial f}{\partial k_1}-\frac{2}{\sqrt{3}}\frac{\partial f}{\partial k_2}\right)-\frac{e\tau E_y}{\hbar}\frac{\partial f}{\partial k_1}+ f = f^0,
\end{split}
\end{equation}
which gives
\begin{equation}
    \frac{e\tau E_x}{\hbar} \left(\frac{1}{\sqrt{3}}\frac{\partial f}{\partial k_1}+\frac{2}{\sqrt{3}}\frac{\partial f}{\partial k_2}\right)-\frac{e\tau E_y}{\hbar}\frac{\partial f}{\partial k_1}+ f = \frac{e\tau}{\hbar}\left(\frac{E_x}{\sqrt{3}}-E_y \right) \frac{\partial f}{\partial k_1} + \frac{2e\tau E_x}{\sqrt{3} \hbar} \frac{\partial f}{\partial k_2}+f= f^0.
\end{equation}
Now, we replace the derivatives by finite differences 
\begin{equation}
\begin{split}
        \frac{\partial f(k_1,k_2)}{\partial k_1} &\approx \frac{f(k_1+\delta k,k_2)-f(k_1-\delta k,k_2)}{2\delta k},\\
        \frac{\partial f(k_1,k_2)}{\partial k_2} &\approx \frac{f(k_1,k_2+\delta k)-f(k_1,k_2-\delta k)}{2\delta k}.
\end{split}
\end{equation}
Now, let us expand $f$ in a Fourier series
\begin{equation}
    f(k_1,k_2) = \sum_{n_1=0}^{\sqrt{\mathcal{N}}-1}\sum_{n_2=0}^{\sqrt{\mathcal{N}}-1} \tilde{f}_{n_1,n_2} e^{i \left(k_1 \hat{g}_1+k_2 \hat{g}_2\right) \cdot \left( n_1 \mathbf{a}_1+ n_2 \mathbf{a}_2 \right)} = \sum_{n_1=0}^{\sqrt{\mathcal{N}}-1}\sum_{n_2=0}^{\sqrt{\mathcal{N}}-1} \tilde{f}_{n_1,n_2} e^{i \frac{\sqrt{3}L}{2}\left(k_1 n_1+k_2 n_2\right) },
\end{equation}
where $\mathbf{a}_1 = L \left(-\frac{1}{2},\frac{\sqrt{3}}{2} \right)$ and $\mathbf{a}_2 = L \left(-1, 0\right).$ We choose $k_1 = \frac{m_1}{\sqrt{\mathcal{N}}}\frac{4\pi}{\sqrt{3}L}$ and $k_2 = \frac{m_2}{\sqrt{\mathcal{N}}}\frac{4\pi}{\sqrt{3}L},$ where $m_1,m_2 \in \left[ 0,\sqrt{\mathcal{N}}-1 \right]$ (recall that $\mathcal{N}$ is defined as the number of unit cells; so $\sqrt{\mathcal{N}}$ is the size along one direction).  This simplifies to
\begin{equation}
    f(m_1,m_2) = \sum_{n_1=0}^{\sqrt{\mathcal{N}}-1}\sum_{n_2=0}^{\sqrt{\mathcal{N}}-1} \tilde{f}_{n_1,n_2} e^{\frac{2 \pi}{\sqrt{\mathcal{N}}} i \left(m_1 n_1+ m_2 n_2\right) }.
\end{equation}
Substituting this into the difference equation, we obtain
\begin{equation}
\begin{split}
    \frac{f(m_1+1,m_2)-f(m_1-1,m_2)}{2 \left(\frac{4\pi}{\sqrt{3}L\sqrt{\mathcal{N}}} \right)} &= \sum_{n_1=0}^{\sqrt{\mathcal{N}}-1}\sum_{n_2=0}^{\sqrt{\mathcal{N}}-1} \tilde{f}_{n_1,n_2} e^{\frac{2 \pi}{\sqrt{\mathcal{N}}} i \left(m_1 n_1+ m_2 n_2\right) } \left[ \frac{i\sin \left(\frac{2\pi}{\sqrt{\mathcal{N}}} n_1\right)}{ \left(\frac{4\pi}{\sqrt{3}L\sqrt{\mathcal{N}}} \right)} \right],    \\
    \frac{f(m_1,m_2+1)-f(m_1,m_2-1)}{2 \left(\frac{4\pi}{\sqrt{3}L\sqrt{\mathcal{N}}} \right)} &= \sum_{n_1=0}^{\sqrt{\mathcal{N}}-1}\sum_{n_2=0}^{\sqrt{\mathcal{N}}-1} \tilde{f}_{n_1,n_2} e^{\frac{2 \pi}{\sqrt{\mathcal{N}}} i \left(m_1 n_1+ m_2 n_2\right) } \left[ \frac{i\sin \left(\frac{2\pi}{\sqrt{\mathcal{N}}} n_2\right)}{ \left(\frac{4\pi}{\sqrt{3}L\sqrt{\mathcal{N}}} \right)} \right]  .  
\end{split}
\end{equation}
Solving for the Fourier coefficients, we find
\begin{equation}
    \tilde{f}_{n_1,n_2}   = \left(\frac{e\tau}{\hbar}\left(\frac{E_x}{\sqrt{3}}-E_y \right)\left[ \frac{i\sin \left(\frac{2\pi}{\sqrt{\mathcal{N}}} n_1\right)}{ \left(\frac{4\pi}{\sqrt{3}L\sqrt{\mathcal{N}}} \right)} \right] + \frac{2e\tau E_x}{\sqrt{3} \hbar} \left[ \frac{i\sin \left(\frac{2\pi}{\sqrt{\mathcal{N}}} n_2\right)}{ \left(\frac{4\pi}{\sqrt{3}L\sqrt{\mathcal{N}}} \right)} \right] + 1\right)^{-1}\tilde{f}^0_{n_1,n_2}.
\end{equation}
As a check of consistency, let $\mathcal{N} \rightarrow \infty,$ then $\frac{i\sin \left(\frac{2\pi}{\sqrt{\mathcal{N}}} n_2\right)}{ \left(\frac{4\pi}{\sqrt{3}L\sqrt{\mathcal{N}}} \right)} \rightarrow  \frac{i\sqrt{3}L n_2}{2}$
\begin{equation}
\begin{split}
    &\frac{e\tau}{\hbar}\left(\frac{E_x}{\sqrt{3}}-E_y \right)\left[ \frac{i\sin \left(\frac{2\pi}{\sqrt{\mathcal{N}}} n_1\right)}{ \left(\frac{4\pi}{\sqrt{3}L\sqrt{\mathcal{N}}} \right)} \right] + \frac{2e\tau E_x}{\sqrt{3} \hbar} \left[ \frac{i\sin \left(\frac{2\pi}{\sqrt{\mathcal{N}}} n_2\right)}{ \left(\frac{4\pi}{\sqrt{3}L\sqrt{\mathcal{N}}} \right)} \right] \\
    &\rightarrow \frac{ie\tau}{\hbar} \left[ L n_1\left( \frac{E_x}{2}- \frac{\sqrt{3}E_y}{2} \right) + E_x L n_2\right]= -\frac{ie \tau}{\hbar} \mathbf{E} \cdot \left(n_1 \mathbf{a}_1+ n_2 \mathbf{a}_2 \right) ,   
\end{split}
\end{equation}
which is expected in the continuum limit.

\section{Twisted Bilayer Graphene on Hexagonal Boron Nitride}

In this section, we compute the band structure of twisted bilayer graphene aligned with hexagonal boron nitride. The Hamiltonian for twisted bilayer graphene is adopted from Ref. \cite{KF18}
\begin{equation}
    \mathcal{H}^\nu = \begin{pmatrix}
        \mathcal{H}_1^\nu  & \mathcal{U}^{\nu{\dagger}} \\
        \mathcal{U}^\nu & \mathcal{H}_2^\nu 
    \end{pmatrix},
\end{equation}
where the intralayer matrix elements are
\begin{equation}
\begin{split}
    \mathcal{H}_1^\nu &= -\hbar v_F \left( \mathcal{R}_+ \mathbf{k}  - \mathbf{K}_\nu\right) \cdot \left( \nu \sigma_x, \sigma_y \right) + m_z \sigma_z, \\
    \mathcal{H}_2^\nu &= -\hbar v_F \left( \mathcal{R}_- \mathbf{k}  - \mathbf{K}_\nu\right) \cdot \left( \nu \sigma_x, \sigma_y \right),
\end{split}
\end{equation}
the interlayer coupling is
\begin{equation}
    \mathcal{U}^\nu = \begin{pmatrix}
        t_\text{AA} & t_\text{AB} \\
        t_\text{AB} & t_\text{AA}
    \end{pmatrix} + \begin{pmatrix}
        t_\text{AA} & t_\text{AB} \alpha^{-\nu}\\
        t_\text{AB}\alpha^{\nu}& t_\text{AA}
    \end{pmatrix} e^{i\nu \mathbf{G}_1}+ \begin{pmatrix}
        t_\text{AA} & t_\text{AB} \alpha^{\nu}\\
        t_\text{AB}\alpha^{-\nu} & t_\text{AA}
    \end{pmatrix}e^{i\nu (\mathbf{G}_1+\mathbf{G}_2)},
\end{equation}
$\alpha = e^{2\pi i/3},$  $\mathbf{K}_\nu = - \frac{4\pi}{3a} \left( \nu, 0\right),$ $\mathbf{G}_1 = \frac{4\pi}{\sqrt{3}L} \left(-\frac{1}{2}, -\frac{\sqrt{3}}{2} \right), $ $\mathbf{G}_2 = \frac{4\pi}{\sqrt{3}L} \left(1,0 \right),$ $L = a/2\sin \left(\theta/2 \right),$ and $\mathcal{R}_\pm$ is the counterclockwise rotation matrix by $\pm \theta/2.$ For pristine twisted bilayer graphene, we use the experimental parameters in Ref. \cite{LLB21}: $a = 2.46$ \AA, $\hbar v_F/a = 2.511$ eV, $t_\text{AA} = 88$ meV, $t_\text{AB} = 110$ meV. We align graphene layer 1 on top of hexagonal boron nitride (hBN) to induce a staggered potential, which gives $m_z = 30$ meV \cite{LPZ22}. Twisted bilayer graphene, without hBN, respects approximately $C_{2z},$ $C_{3z},$ $C_{2x},$ $C_{2y}$ rotation symmetries and time-reversal $\mathcal{T}$ symmetry. In this setting, each valley hosts two Dirac cones at the mini-zone corners due to $\mathcal{T}C_{2z}.$ However, in the presence of a staggered potential on one of the layers, $C_{2z}$ symmetry is broken because of sublattice inequivalence, and $C_{2y}$ and $C_{2x}$ symmetries are broken due to layer inequivalence. This allows the Dirac cones to gap out, a process that generically endows the resulting singlet bands with nontrivial topology. In this model, since most parameters have been fixed to their experimental values, there is only one free parameter, the twist angle $\theta.$

\begin{figure}
    \centering
    \includegraphics[width=4in]{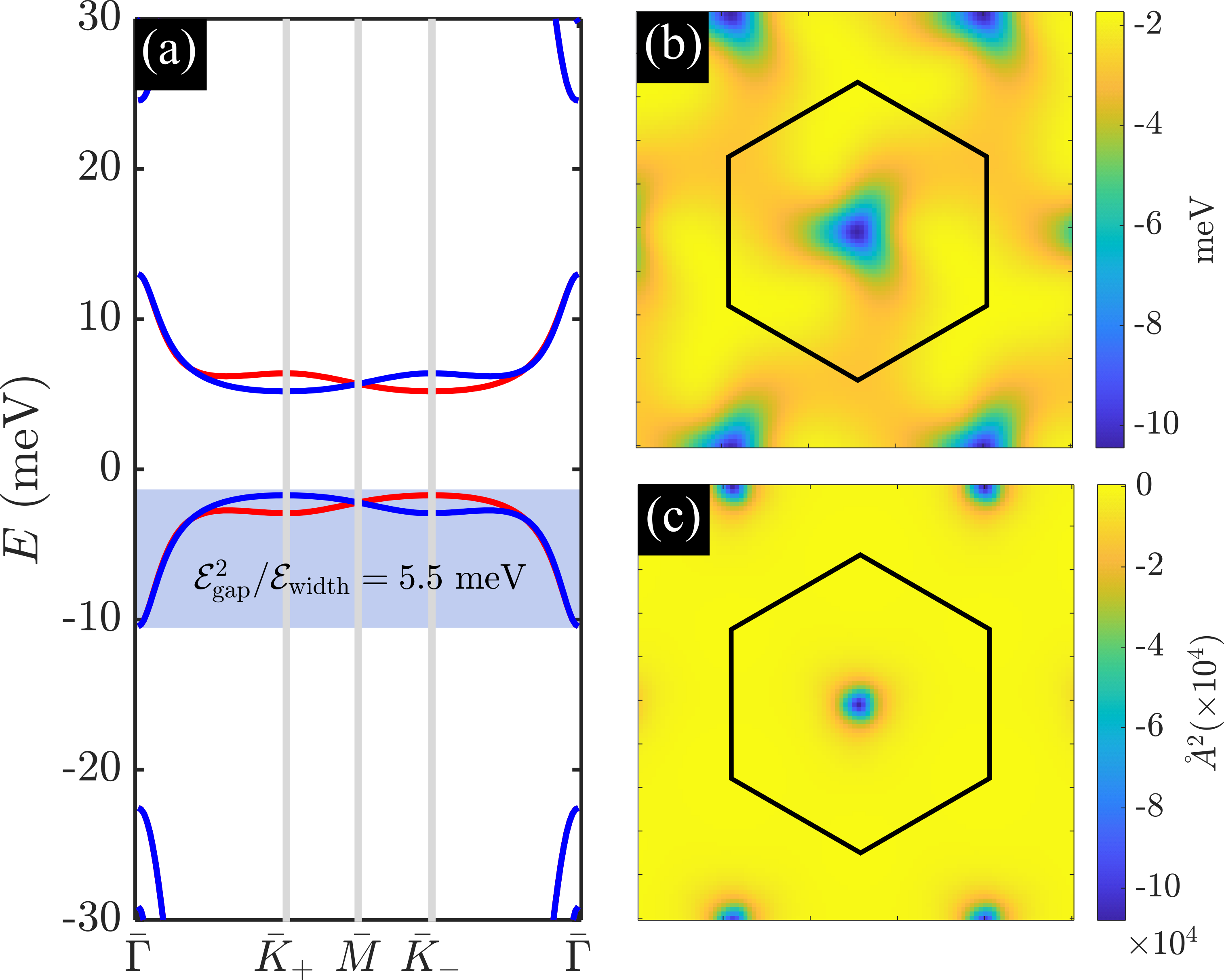}
    \caption{(a) Band structure of twisted bilayer graphene on top of hexagonal boron nitride which induces a mass term $m_z = 30$ meV on the bottom  graphene layer. The two bands, per valley, near charge neutrality have Chern numbers $\pm 1.$ The blue and red bands are from valley $\nu = -1$ and $\nu = +1$ respectively. The band dispersion and Berry curvature for the valence band in valley $\nu = +1$ are shown in (b) and (c) respectively. }
    \label{fig:TBG band structure}
\end{figure}

We find that near $\theta = 1.0^\circ,$ the bands near charge neutrality are quite flat, as shown in Fig. \ref{fig:TBG band structure}. Focusing on the valence band for the moment, we find the bandwidth to be about $ \mathcal{E}_\text{width} \approx 8.7$ meV and the band gap to be about $ \mathcal{E}_\text{gap} \approx 6.9$ meV, giving a ratio $\mathcal{E}_\text{gap}^2/\mathcal{E}_\text{width} \approx 5.5$ meV. The band has Chern number $\mathcal{C} = -\nu.$ In this regime, $L = a/(2\sin(\theta/2)) \approx 140$ \AA. So the required field strength to crossover to the strong-coupling limit for a system with scattering time $\tau = 1$ ps is $E = \hbar/ eL\tau \approx 0.5$ kV/cm. The energy scale of this field strength is $eEL \approx 0.7$ meV, which is much smaller than the field strength needed to activate significant Zener-Landau tunneling. Based on this analysis, we conclude that twisted bilayer graphene aligned with hexagonal boron nitride is a promising candidate to observe geometric oscillations.

\end{document}